\documentclass[review,12pt,authoryear]{elsarticle}

\usepackage{amssymb}
\usepackage{amsthm}
\usepackage{framed} 
\usepackage{amsmath,color}
\usepackage{mathrsfs}
\usepackage{graphicx}
\usepackage{epstopdf}
\usepackage{float}
\usepackage{caption}
\usepackage{subcaption}
\usepackage{bm}
\usepackage{bbm}
\usepackage{upgreek} 
\usepackage{mathrsfs}
\usepackage{cleveref}
\usepackage{soul}
\usepackage{accents}

\usepackage{tikz}
\usetikzlibrary{patterns}
\usetikzlibrary{arrows.meta}

\usepackage{color,soul} 
\usepackage{color} 
\usepackage{bm}
\biboptions{sort&compress}
\soulregister\citep7 
\soulregister\citet7 
\soulregister\citealp7 
\newsavebox{\measurebox} 
\usepackage{titlesec} 
\usepackage[T1]{fontenc} 
\usepackage{lmodern} 
\input{glyphtounicode}  

\newcommand{\tm}{\textrm} 

\def\onedot{$\mathsurround0pt\ldotp$}
\def\cddot{
  \mathbin{\vcenter{\baselineskip.67ex
    \hbox{\onedot}\hbox{\onedot}}%
  }}%
\def\cdddot#1{
  \mathbin{\vcenter{\baselineskip.67ex
    \hbox{\onedot}\hbox{\onedot}\hbox{\onedot}%
  }}%
}

\journal{Journal of the Mechanics and Physics of Solids}

\makeatletter
\def\@author#1{\g@addto@macro\elsauthors{\normalsize%
    \def\baselinestretch{1}%
    \upshape\authorsep#1\unskip\textsuperscript{%
      \ifx\@fnmark\@empty\else\unskip\sep\@fnmark\let\sep=,\fi
      \ifx\@corref\@empty\else\unskip\sep\@corref\let\sep=,\fi
      }%
    \def\authorsep{\unskip,\space}%
    \global\let\@fnmark\@empty
    \global\let\@corref\@empty  
    \global\let\sep\@empty}%
    \@eadauthor={#1}
}
\makeatother

\titleformat{\paragraph}
{\normalfont\normalsize\itshape}{\theparagraph}{1em}{}
\titlespacing*{\paragraph}
{0pt}{3.25ex plus 1ex minus .2ex}{1.5ex plus .2ex}

\begin{document}

\begin{frontmatter}



\title{A phase field model for elastic-gradient-plastic solids undergoing hydrogen embrittlement}


\author{Philip K. Kristensen \fnref{DTU}}

\author{Christian F. Niordson\fnref{DTU}}

\author{Emilio Mart\'{\i}nez-Pa\~neda\corref{cor1}\fnref{IC}}
\ead{e.martinez-paneda@imperial.ac.uk}

\address[DTU]{Department of Mechanical Engineering, Technical University of Denmark, DK-2800 Kgs. Lyngby, Denmark}

\address[IC]{Department of Civil and Environmental Engineering, Imperial College London, London SW7 2AZ, UK}

\cortext[cor1]{Corresponding author.}

\begin{abstract}
We present a gradient-based theoretical framework for predicting hydrogen assisted fracture in elastic-plastic solids. The novelty of the model lies in the combination of: (i) stress-assisted diffusion of solute species, (ii) strain gradient plasticity, and (iii) a hydrogen-sensitive phase field fracture formulation, inspired by first principles calculations. The theoretical model is numerically implemented using a mixed finite element formulation and several boundary value problems are addressed to gain physical insight and showcase model predictions. The results reveal the critical role of plastic strain gradients in rationalising decohesion-based arguments and capturing the transition to brittle fracture observed in hydrogen-rich environments. Large crack tip stresses are predicted, which in turn raise the hydrogen concentration and reduce the fracture energy. The computation of the steady state fracture toughness as a function of the cohesive strength shows that cleavage fracture can be predicted in otherwise ductile metals using sensible values for the material parameters and the hydrogen concentration. In addition, we compute crack growth resistance curves in a wide variety of scenarios and demonstrate that the model can appropriately capture the sensitivity to: the plastic length scales, the fracture length scale, the loading rate and the hydrogen concentration. Model predictions are also compared with fracture experiments on a modern ultra-high strength steel, AerMet100. A promising agreement is observed with experimental measurements of threshold stress intensity factor $K_{th}$ over a wide range of applied potentials.
\end{abstract}


\end{frontmatter}



\section{Introduction}
\label{Sec:Introduction}

Variational phase field models for fracture are receiving much attention due to their modeling capabilities (see \citealp{Wu2020} for a review). The phase field framework enables predicting advanced fracture features without remeshing, such as crack nucleation at arbitrary sites, crack growth along complex trajectories, and branching and coalescence of multiple cracks (\citealp{Borden2012}; \citealp{McAuliffe2016}; \citealp{TAFM2020}). These predictions are based on the energy balance first proposed by \citet{Griffith1920}, with fracture occurring when the energy release rate of system reaches a critical value, $G_c$. In addition, both the discrete crack phenomenon and damage, in a continuum sense, can be captured in the phase field framework \citep{Francfort1998,Pham2011}. Since the pioneering numerical experiments by \citet{Bourdin2000}, phase field fracture models have gained increased interest. Recent applications include hydraulic fracturing (\citealp{Mikelic2015}; \citealp{Cajuhi2018}), ductile damage (\citealp{Borden2016}; \citealp{Alessi2018}), lithium-ion batteries \citep{Miehe2015,Zhao2016}, composites delamination \citep{Carollo2017,Quintanas-Corominas2019}, rock fracture \citep{Zhou2018} and functionally graded materials \citep{CPB2019}, \textit{inter alia}.\\

Recently, the success of phase field fracture methods has also been extended to model the phenomenon of hydrogen embrittlement (\citealp{CMAME2018}; \citealp{Duda2018}; \citealp{Anand2019}; \citealp{Wu2020b}). Hydrogen severely degrades the ductility and the fracture resistance of metals, with the fracture toughness of modern steels decreasing by up to 90\% \citep{Gangloff2003}. The problem is now pervasive in the transport, energy, construction and defence sectors due to the ubiquity of hydrogen and the higher susceptibility of high strength alloys \citep{Gangloff2012}. Hydrogen atoms enter the material, diffuse through the crystal lattice and are attracted to regions of high hydrostatic stress, where damage occurs through mechanisms that are still being debated \citep{Robertson2015,Tehranchi2019,Lynch2019,Yu2019,Harris2018,Shishvan2020}. By accounting for the degradation of the fracture energy with hydrogen content, multi-physics phase field fracture models capture the trends shown in the experiments (see, e.g., \citealp{CS2020}), while establishing a computational framework capable of dealing with the complex scenarios relevant to engineering practice. Phase field models are bringing a paradigm change to the hydrogen assisted cracking community, where modeling efforts were focused on discrete methods. Cohesive zone models have particularly enjoyed great popularity and have proven capable of capturing the strength degradation with increasing hydrogen content (\citealp{Serebrinsky2004}; \citealp{Scheider2008}; \citealp{Moriconi2014}; \citealp{EFM2017}; \citealp{Yu2017}). However, discrete methods are limited when dealing with complex fracture conditions. Moreover, conventional continuum models fail to resolve the critical length scale of hydrogen assisted fracture. Cracking occurs very close to the crack tip, at 1 $\mu$m or less \citep{Gangloff2003a}, where dislocation-based hardening governs material behavior. Large gradients of plastic strain are present within microns ahead of the crack tip, requiring a significant storage of \emph{geometrically necessary} dislocations (GNDs) to accommodate lattice curvature \citep{Ashby1970,IJSS2015}.
The increased dislocation density associated with large gradients of plastic deformation promotes strain hardening and leads to crack tip stresses that are much larger than those predicted by conventional plasticity. The flow strength elevation associated with plastic strain gradients has been quantified in a wide range of experiments, from wire torsion \citep{Fleck1994} to indentation \citep{Nix1998}; see \citep{Voyiadjis2019} for a review. Continuum models can be enriched to capture the local strengthening observed when the macroscopic strain field varies over microns. In this regard, the development of phenomenological strain gradient plasticity (SGP) theories has received particular attention (\citealp{Dillon1970}; \citealp{Gao1999}; \citealp{Fleck2001}; \citealp{Anand2005}). SGP models have been used to investigate the influence of plastic strain gradients ahead of stationary and propagating cracks (\citealp{Wei1997}; \citealp{Komaragiri2008}; \citealp{IJP2016}; \citealp{Seiler2016}). Predictions show notable strain gradient hardening, with crack tip stresses being substantially higher that those predicted by conventional plasticity. A similar level of local crack tip strengthening to that predicted by SGP is also found in discrete dislocation dynamics simulations \citep{Balint2005,Chakravarthy2010}. The impact of this stress elevation on the understanding and modeling of hydrogen embrittlement is twofold. First, given the dependence of the hydrogen content on the hydrostatic stress, a high hydrogen concentration is attained close to the crack tip surface, in agreement with neutron activation and SIMS measurements \citep{Gerberich2012,Mao1998}. Secondly, large crack tip tensile stresses and hydrogen concentrations rationalise decohesion-based mechanisms on high strength alloys \citep{AM2016}. However, a continuum modeling framework capable of explicitly predicting cracking while accounting for the dislocation hardening mechanisms governing crack tip deformation has not been presented yet.\\

In this work, we aim at presenting a computationally compelling framework for modeling fracture in embrittled alloys, which is informed by atomistic and micromechanical considerations. The model builds upon: (i) higher order strain gradient plasticity (\citealp{Gudmundson2004}; \citealp{Fleck2009}), to accurately characterise crack tip stresses; (ii) a coupled mechanical and hydrogen diffusion response, driven by chemical potential gradients (\citealp{Sofronis1989}; \citealp{Diaz2016b}); (iii) a phase field description of fracture \citep{Miehe2010a}; and (iv) a hydrogen-dependent fracture energy degradation law, grounded on first principles calculations \citep{Jiang2004a}. We demonstrate the potential of the proposed modeling framework in (1) providing relevant physical insight, by predicting the ductile-to-brittle transition observed when incorporating hydrogen, and (2) quantitatively capturing experimental measurements across a wide range of potentials. The remainder of this manuscript is organized as follows. The theoretical framework is presented in Section \ref{Sec:Theory}. The finite element implementation is briefly described in Section \ref{Sec:Numerical}, with further details provided in \ref{Sec:AppendixFEMDetails}. Representative results are shown in Section \ref{Sec:Results}. First, we aim at gaining physical insight into model predictions by computing crack tip stresses and crack growth resistance curves for a wide range of scenarios. Secondly, stress intensity factor thresholds are predicted as a function of the applied potential for AerMet100, so as to benchmark the capabilities of the model in quantitatively reproducing experiments. Finally, concluding remarks are given in Section \ref{Sec:Conclusions}.\\ 

\noindent \textit{Notation.}
We use lightface italic letters for scalars, e.g. $\phi$, upright bold letters for vectors, e.g. $\mathbf{u}$, and bold italic letters, such as $\bm{\sigma}$, for second and higher order tensors. Inner products are denoted by a number of vertically stacked dots, corresponding to the number of indices over which summation takes place, such that $\bm{\sigma}\cddot\bm{\varepsilon} = \sigma_{ij}\varepsilon_{ij}$, with indices referring to a Cartesian coordinate system. The full inner product of a tensor with itself is denoted $|\bm{\tau}|^2=\bm{\tau} \cdddot{} \bm{\tau} =\tau_{ijk}\tau_{ijk}$. The gradient and the Laplacian are respectively denoted by $\nabla\mathbf{u}= u_{i,j}$ and $\Delta\phi=\phi_{,ii}$. Finally, divergence is denoted by $\nabla\cdot\bm{\sigma}=\sigma_{ij,j}$, the trace of a second order tensor is written as $\tm{tr} \,\bm{\varepsilon}=\varepsilon_{ii}$, and the deviatoric part of a tensor is written as $\bm{\sigma}'=\sigma_{ij}-\delta_{ij} \sigma_{kk}$, with $\delta_{ij}$ denoting the Kronecker delta. 

\section{Theory}
\label{Sec:Theory}

In this section, we formulate our theory, which couples deformation, damage and hydrogen transport in elastic-plastic bodies. The theory refers to the response of a solid occupying an arbitrary domain $\Omega \subset {\rm I\!R}^n$ $(n \in[1,2,3])$, with external boundary $\partial \Omega\subset {\rm I\!R}^{n-1}$, on which the outwards unit normal is denoted as $\mathbf{n}$.

\subsection{Kinematics}

The primal kinematic variables of the model are the displacement field $\mathbf{u}$, the plastic strain tensor $\bm{\varepsilon}^p$, the damage phase field $\phi$, and the hydrogen concentration $C$. We restrict our attention to small strains and isothermal conditions. Accordingly, the strain tensor $\bm{\varepsilon}$ is given by
\begin{equation}
    \bm{\varepsilon} = \frac{1}{2}\left(\nabla\mathbf{u}^T+\nabla\mathbf{u}\right),
\end{equation}

\noindent and we adopt the standard partition of strains into elastic and plastic components: $\bm{\varepsilon} = \bm{\varepsilon}^e + \bm{\varepsilon}^p$.\\

Regarding fracture, a smooth continuous scalar function, $\phi \in [0;1]$, is introduced, which describes the degree of damage in a given material point in $\Omega$. This function will be referred to as the \textit{phase field}. The phase field takes the value $0$ when the material is intact and $1$ when the material is fully broken. Since $\phi$ is smooth and continuous, discrete cracks are represented in a diffuse fashion. The smearing of cracks is controlled by a phase field length scale $\ell$. The purpose of this diffuse representation is to introduce the following approximation of the fracture energy over a discontinuous surface $\Gamma$: 
\begin{equation}
    \Psi^s=\int_{\Gamma} G_c \, \text{d}S \approx \int_\Omega G_c\gamma(\phi,\nabla\phi) \, \text{d}V, \hspace{1cm} \tm{for } \ell\rightarrow 0,
\end{equation}

\noindent where $\gamma$ is the crack surface density functional and $G_c$ is the critical energy release rate \citep{Griffith1920,Irwin1956}. This approximation circumvents the need to track discrete crack surfaces, which is a major complication in numerical fracture models.\\

Regarding the diffusion of solute species. The concentration of hydrogen in a material point in $\Omega$ is given by the continuous smooth scalar function $C$. Due to conservation of mass, the rate of change in time of the hydrogen concentration $\Dot{C}=dC/dt$ is equal to the concentration flux $\mathbf{J}\cdot\mathbf{n}=-\rho $ through the boundary $\partial\Omega$: 
\begin{equation}\label{eq:weakDiffusion}
    \int_\Omega \dot{C} \, \text{d}V  +\int_{\partial\Omega}\mathbf{J}\cdot\mathbf{n} \, \text{d}S =0,
\end{equation}

\noindent with $\rho$ denoting the inwards boundary flux. Alternatively, since the above must hold for any volume and by use of Gauss' divergence theorem, Eq. (\ref{eq:weakDiffusion}) can be formulated in a point-wise manner as: 
\begin{equation}
    \dot{C} + \nabla\cdot \mathbf{J}=0.\label{eq:Hydrogen_strong}
\end{equation}

The diffusion of solute species is driven by the chemical potential $\mu$, which is also a smooth scalar function in $\Omega$. The flux $\mathbf{J}$ is related to the chemical potential $\mu$ through a linear Onsager relation \citep{Kirchheim2004}
\begin{equation}
    \mathbf{J}=-\dfrac{DC}{RT}\nabla\mu, \label{eq:FricksLaw}
\end{equation}

\noindent where $D$ is the diffusion coefficient of the material, $R$ is the gas constant and $T$ is the absolute temperature. 

\subsection{Principle of virtual work. Balance of forces}

\begin{figure}
    \centering
    \includegraphics[width=0.7\linewidth]{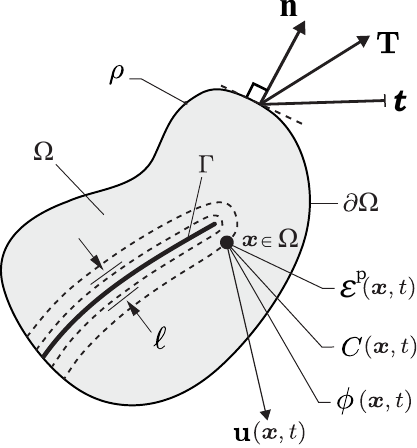}
    \caption{Schematic representation of a solid body $\Omega$, with a sharp crack $\Gamma$ represented diffusively through the phase field.}
    \label{fig:potato}
\end{figure}
The balance equations for the coupled system are now derived using the principle of virtual work. Consider the four-field boundary value problem outlined in Fig. \ref{fig:potato}. The Cauchy stress $\bm{\sigma}$ is introduced, which is work conjugate to the elastic strains $\bm{\varepsilon}^e$. Correspondingly, for an outwards unit normal $\mathbf{n}$ on the boundary $\partial\Omega$ of the solid, a traction $\mathbf{T}$ is defined, which is work conjugate to the displacements $\mathbf{u}$. The plastic response is given by the so-called micro-stress tensor $\bm{q}$, work conjugate to the plastic strain $\bm{\varepsilon}^p$, and the higher order stress tensor $\bm{\tau}$, work conjugate to the plastic strain gradient $\nabla\bm{\varepsilon}^p$. A higher order traction $\bm{t}$ is also introduced on the boundary of the solid as work conjugate to the plastic strains. Regarding damage, we introduce a scalar stress-like quantity $\omega$, which is work conjugate to the phase field $\phi$, and a phase field micro-stress vector $\bm{\upxi}$ that is work conjugate to the gradient of the phase field $\nabla\phi$. The phase field is assumed to be driven by the displacement problem alone. As a result, no external traction is associated with $\phi$. Lastly, a boundary flux of hydrogen $\rho$ is defined as work conjugate of the chemical potential in the diffusion problem. Accordingly, in the absence of body forces, the external virtual work is given by: 
\begin{equation}\label{eq:ExternalVirtualWork}
    \delta W_{ext} = \int_{ \partial\Omega} \big\{ \mathbf{T}\cdot\delta \mathbf{u} + \bm{t}\cddot\delta\bm{\varepsilon}^p+\rho \, \delta \mu \big\} \, \text{d}S,
\end{equation}

\noindent where $\delta$ denotes a virtual quantity. The corresponding internal work reads: 
\begin{align}\label{eq:InternalVirtualWork}
    \delta W_{int} = \int_\Omega \big\{ \bm{\sigma}\cddot\delta\bm{\varepsilon} &+\left(\bm{q}-\bm{\sigma}'\right)\cddot \delta \bm{\varepsilon}^p + \bm{\tau} \cdddot \,\delta\nabla \bm{\varepsilon}^p  + \omega\delta\phi\nonumber \\ 
    &+\bm{\upxi}\cdot\delta\nabla\phi
    +\dot{C}\delta\mu-\mathbf{J} \cdot \delta \nabla \mu \big\} \, \text{d}V,
\end{align}

\noindent where $\bm{\sigma}'$ denotes the deviatoric part of $\bm{\sigma}$. For simplicity, the prime symbol is omitted from $\bm{q}$ and $\bm{\tau}$, as they can be inherently defined to be deviatoric. Eqs. (\ref{eq:ExternalVirtualWork})-(\ref{eq:InternalVirtualWork}) must hold for an arbitrary domain $\Omega$ and for any kinematically admissible variations of the virtual quantities. Thus, by making use of the fundamental lemma of calculus of variations, the local force balances are given by: 
\begin{equation}
    \begin{split}
        &\nabla\cdot\bm{\sigma}=0  \\
        &\nabla\cdot\bm\tau + \bm{\sigma}'-\bm{q} =0 \\ 
        &\nabla\cdot\bm{\upxi}-\omega =0 \\
        &\dot{C} + \nabla\cdot \mathbf{J}=0
    \end{split}\hspace{2cm} \tm{in } \Omega,\label{eq:balance}
\end{equation}
\noindent with natural boundary conditions: 
\begin{equation}
    \begin{split}
        \bm{\sigma}\cdot\mathbf{n}=\mathbf{T} \\
        \bm{\tau}\cdot\mathbf{n} = \bm{t} \\
         \bm{\upxi} \cdot \mathbf{n}=0 \\
        -\mathbf{J}\cdot\mathbf{n}=\rho
    \end{split} \hspace{2cm} \tm{on } \partial\Omega.\label{eq:balance_BC}
\end{equation}

\subsection{Energy imbalance}
The first and second law of thermodynamics can be expressed through the Helmholtz free energy per unit volume $\Psi\left(\bm{\varepsilon},\nabla\bm{\varepsilon}^p,\phi,\nabla\phi,C\right)$ and the external work $W_{ext}$ in the Clausius-Duhem inequality: 
\begin{equation}
    \int_\Omega \dot{\Psi} \, \text{d}V - \int_{\partial\Omega}\dot{W}_{ext} \, \text{d}S \leq 0.
\end{equation}

Inserting Eqs. (\ref{eq:balance})-(\ref{eq:balance_BC}) and applying the divergence theorem, the inequality may be restated as: 
\begin{equation}
\begin{split}
        \int_\Omega \dot{\Psi} \, \text{d}V - \int_{\Omega}{\big\{ }\bm{\sigma}\cddot\nabla\dot{\mathbf{u}} &+  \left(\bm{q}-\bm{\sigma}'\right)\cddot\dot{\bm{\varepsilon}}^p+\bm{\tau}\cdddot \nabla \nabla\dot{\bm{\varepsilon}}^p\\
 &+ \omega\dot{\phi}+\bm{\upxi}\cdot \nabla\dot{\phi} +\dot{C}\, \dot{\mu} - \mathbf{J}\cdot\nabla\dot{\mu}  \big\} \, \text{d}V \leq 0. \end{split}
\end{equation}

Since the above must hold for any volume $\Omega$, it follows that it must also hold in a local fashion, such that: 
\begin{align}\label{eq:Energy_balance}
    \left( \bm{\sigma}- \dfrac{\partial\Psi}{\partial\bm{\varepsilon}^e} \right) & \cddot\dot{\bm{\varepsilon}}^e  + \left(\bm{q}-\dfrac{\partial\Psi}{\partial\bm{\varepsilon}^p}\right)\cddot\dot{\bm{\varepsilon}}^p+\left(\bm{\tau}-\dfrac{\partial\Psi}{\partial\nabla\bm{\varepsilon}^p}\right)\cdddot \nabla\nabla\dot{\bm{\varepsilon}}^p +\left(\omega-\dfrac{\partial\Psi}{\partial\phi}\right)\dot{\phi} \nonumber \\
     & + \left( \bm{\upxi}-\dfrac{\partial\Psi}{\partial\nabla\phi} \right) \cdot\nabla\dot{\phi} + \left[ \left( \mu-\dfrac{\partial\Psi}{\partial C} \right) \dot{C} - \mathbf{J} \cdot \nabla \mu \right] \geq 0.
\end{align}

\noindent To satisfy this inequality, a free energy function $\Psi$ is proposed, which is composed by the chemo-elastic energy stored in the bulk  $\Psi^e (\bm{\varepsilon}^e, \, C, \, \phi)$, the plastic defect energy $\Psi^p (\nabla \bm{\varepsilon}^p)$, the crack surface energy $\Psi^s (\phi, \, \nabla \phi, \, C)$, and the chemical free energy $\Psi^c (C)$. Thus, consider a solid with bulk modulus $K$, shear modulus $Q$, number of lattice sites $N$ (with a lattice site occupancy $\theta_L=C/N$), and partial molar volume of hydrogen $\overline{V}_H$. Denoting the reference chemical potential and hydrogen concentration as, respectively, $\mu^0$ and $C^0$, the free energy is defined as:
\begin{align} \label{eq:Free_energy}
  \Psi =& \underbrace{ (1 - \phi)^2 \psi^e -K\overline{V}_H\left(C -C^0\right)\tm{tr}\bm{\varepsilon}^e}_{\Psi^e} + \underbrace{Q L_E^2 \nabla \bm{\varepsilon}^p}_{\Psi^p}+ \underbrace{\dfrac{1}{2} G_c (C) \left(\dfrac{1}{\ell}\phi^2+\ell|\nabla\phi|^2\right)}_{\Psi^s} \nonumber \\
         &+\underbrace{\mu^0 C + RTN\left[\theta_L\ln{\theta_L}+(1-\theta_L)\ln{(1-\theta_L)}\right]}_{\Psi^c}.
\end{align}

\noindent Here, $\psi^e$ is the elastic strain energy density, which constitutes the driving force for fracture. The length scale $L_E$ quantifies the degree to which the material exhibits energetic gradient hardening; for example, due to long range back-stresses associated with the stored elastic energy of GNDs. Also, we emphasize that the fracture resistance of the material, in terms of the critical energy release rate, is defined as a function of the hydrogen concentration $G_c (C)$.

\subsection{Constitutive relations}

Consistent with the free energy (\ref{eq:Free_energy}), we proceed now to develop a constitutive theory that couples the four primary kinematic variables of the problem.

\subsubsection{Chemo-elasticity}

Following a continuum damage mechanics approach, the phase field damage variable $\phi$ degrades the elastic stiffness of the solid. The degradation function is assumed to be of quadratic form:
\begin{equation}\label{eq:degfunction}
    g(\phi) = (1-\phi)^2,
\end{equation}

\noindent and the elastic strain energy density $\psi^e$ is defined as a function of the elastic strains $\bm{\varepsilon}^e$ and the isotropic elastic stiffness tensor $\bm{\mathcal{L}}_0$ in the usual manner:
\begin{equation}
   \psi^e=\dfrac{1}{2}\bm{\varepsilon}^e\cddot \mathbf{\mathcal{L}}_0\cddot\bm{\varepsilon}^e.
\end{equation}

The Cauchy stress tensor $\bm{\sigma}$ follows immediately from the free energy definition (\ref{eq:Free_energy}) as: 
\begin{equation}\label{eq:CauchyStress}
    \bm{\sigma} = \dfrac{\partial \Psi}{\partial \bm{\varepsilon}^e} = (1-\phi)^2\bm{\mathcal{L}}_0\cddot\bm{\varepsilon}^e-K\overline{V}_H\left(C -C_0\right)\mathbf{I}
\end{equation}

\noindent with the second term, which involves the lattice dilation, being omitted in hydrogen embrittlement analyses due to its negligible influence \citep{Hirth1980}.

\subsubsection{Strain gradient plasticity}

We consider higher order strain gradient plasticity, incorporating both dissipative and energetic strain gradient contributions \citep{Gudmundson2004,JMPS2019}. Thus, both the micro-stress tensor $\bm{q}$ and the higher order stress tensor $\bm{\tau}$ can be additively decomposed into their energetic and dissipative parts:
\begin{equation}
\bm{q} = \bm{q}^D + \bm{q}^E, \hspace{1cm} \bm{\tau} = \bm{\tau}^D+\bm{\tau}^E.
\end{equation} 

Consistent with our free energy definition (\ref{eq:Free_energy}), plastic deformation is assumed to be a purely dissipative process: $\bm{q}^E=\partial\Psi/\partial\bm{\varepsilon}^p=0$. Conversely, both energetic and dissipative terms are considered in relation to the plastic strain gradients. Accordingly, the plastic dissipation rate reads:
\begin{equation}
    \dot{w}^p =\bm{q} \cddot \dot{\bm{\varepsilon}}^p + \bm{\tau}^D \cdddot{} \nabla\dot{\bm{\varepsilon}}^p,
\end{equation}

\noindent where $\dot{w}^p (\dot{E}^p)$ is given in terms of a combined effective plastic rate:
\begin{equation}
\dot{E}^p = \left(\frac{2}{3}|\dot{\bm{\varepsilon}}^p|^2 + L_D^{2}|\nabla\dot{\bm{\varepsilon}}^p|^2 \right)^{1/2}.
\end{equation}

\noindent Here, $L_D$ is the dissipative length scale, which quantifies the degree to which the material exhibits dissipative strengthening; for example, via mechanisms such as forest hardening. A thermodynamically consistent framework is obtained by defining an effective stress $\Sigma=\partial \dot{w}^p/ \partial \dot{E}^p$, work conjugate to $\dot{E}^p$. The constitutive definitions of the dissipative stresses readily follow:
\begin{equation}\label{eq:qDandTauD}
\bm{q} = \frac{\partial \dot{w}^p}{\partial \dot{\bm{\varepsilon}}^p} = \dfrac{2}{3}\dfrac{\Sigma}{\dot{E}^p}\dot{\bm{\varepsilon}}^p, \hspace{1cm} \bm{\tau}^D = \frac{\partial \dot{w}^p}{\partial \nabla \dot{\bm{\varepsilon}}^p} =L_D^2\dfrac{\Sigma}{\dot{E}^p}\nabla\dot{\bm{\varepsilon}}^p.
\end{equation}

On the other hand, the energetic part of the higher order stress is derived from the free energy definition (\ref{eq:Free_energy}) as
\begin{equation}\label{eq:tauE}
   \bm{\tau}^E = \dfrac{\partial\Psi}{\partial\nabla\bm{\varepsilon}^p}=Q L_E^2 \nabla \bm{\varepsilon}^p. 
\end{equation}

For simplicity, we choose to define a single reference plastic length scale $L_p=L_E=L_D$, although the individual contributions from energetic and dissipative higher order gradients will also be explored.\\

The displacement $\mathbf{u}$ and plastic strain $\bm{\varepsilon}^p$ solutions are coupled by the deviatoric Cauchy stress, as evident from (\ref{eq:balance})b, and by the total strain decomposition. Given that the tensile response prior to fracture is typically unaffected by hydrogen, no explicit coupling between the hydrogen content and plasticity is defined. Lastly, as discussed below, fracture is assumed to be driven by the elastic strain energy density.

\subsubsection{Phase field fracture}

The constitutive relations for the micro-stress variables work conjugate to the phase field and the phase field gradient are obtained from the free energy (\ref{eq:Free_energy}). Thus, the scalar microstress $\omega$ is given by:
\begin{equation}\label{eq:consOmega}
    \omega = \dfrac{\partial\Psi}{\partial\phi} = -2(1-\phi)\psi^e+G_c (C) \dfrac{\phi}{\ell}.
\end{equation}

\noindent Similarly, the phase field microstress vector $\bm{\upxi}$ reads:
\begin{equation}\label{eq:consXi}
    \bm{\upxi} = \dfrac{\partial\Psi}{\partial\nabla\phi} = G_c (C) \, \ell \, \nabla\phi.
\end{equation}

Now, insert (\ref{eq:consOmega}) and (\ref{eq:consXi}) into the phase field local balance (\ref{eq:balance}c). Neglecting the concentration gradient along the small region where $\nabla \phi \neq 0$, the local force balance can be reformulated as:
\begin{equation}\label{eq:PhaseFieldStrongForm}
  G_c ( C) \left( \frac{\phi}{\ell} - \ell \Delta \phi \right) - 2 (1 - \phi) \psi^e = 0  
\end{equation}

As evident from (\ref{eq:PhaseFieldStrongForm}), fracture in the elastic-plastic solid is driven solely by the elastic component of the material strain energy density. The same assumption was adopted by \citet{Duda2015}. The plastic contribution may also be weighted differently, through an \textit{ad hoc} degradation function. These and other possibilities, including defining an explicit relation between the plastic yield condition and the damage variable, have been explored in the realm of phase field modeling of ductile fracture, see \citep{Alessi2018}. In addition, note that the coupling with the diffusion problem takes place through the fracture energy dependency to the hydrogen content. The specific choice of the function $G_c(C)$ is inspired by first principles, as discussed below.

\subsubsection{Hydrogen transport}

The gradient of the chemical potential $\nabla \mu$ is the driving force for hydrogen diffusion. The constitutive relation for $\mu$ can be determined from the free energy definition (\ref{eq:Free_energy}) as: 
\begin{equation}\label{eq:constichempot}
    \mu= \dfrac{\partial\Psi}{\partial C} = \mu_0 + RT\ln{\dfrac{\theta_L}{1-\theta_L}}- \overline{V}_H \sigma_H + \frac{1}{2} \dfrac{dG_c (C)}{dC}\left(\dfrac{\phi^2}{\ell}+\ell|\nabla\phi|^2\right).
\end{equation}

As evident from (\ref{eq:constichempot}) and (\ref{eq:FricksLaw}), hydrogen atoms diffuse from regions of high chemical potential to regions of low chemical potential. Hydrogen transport is enhanced by lattice dilatation, as characterised by hydrostatic tensile stresses $\sigma_H$. Note that, as opposed to the choice made in \citep{CMAME2018}, the stress-dependent part of $\mu$ is chosen to be subjected to the degradation function $g(\phi)$. In addition, the last term in (\ref{eq:constichempot}) enhances hydrogen transport from damaged regions to pristine regions. However, as discussed in \citep{CMAME2018}, the definition of sound chemical boundary conditions in the presence of a propagating crack requires careful consideration. As elaborated in Section \ref{subsec:ChemBC}, we choose to neglect the last term in (\ref{eq:constichempot}) and implement a penalty-based \emph{moving} chemical boundary condition to capture how the environment promptly occupies the space created with crack advance. Accordingly, the constitutive equation for the hydrogen flux can be readily obtained by considering (\ref{eq:FricksLaw}). Thus, after adopting the common assumptions of low occupancy ($\theta_L<< 1$) and constant interestitial sites concentration ($\nabla N=0$), the flux reads:
\begin{equation}\label{eq:constiflux}
    \mathbf{J} = -D\nabla C + \dfrac{DC}{RT}\overline{V}_H\nabla\sigma_H.
\end{equation}

The relation between the fracture energy and the hydrogen content remains to be defined. In an implicit multi-scale approach, we define $G_c$ according to the surface energy degradation with hydrogen coverage obtained from quantum mechanical calculations. The aim is to predict the sensitivity of the macroscopic fracture energy to hydrogen by quantifying the reduction in the atomic bond energy, without resorting to empirical parameters. The choice is inspired by the work of \citet{Serebrinsky2004} in the context of cohesive zone models. Density Functional Theory (DFT) calculations show that the atomic decohesion strength depends sensitively on the hydrogen surface coverage along atomic planes (\citealp{VanderVen2003}; \citealp{Jiang2004a}; \citealp{Kirchheim2015}). Based on the recent first principles calculations by \citet{Alvaro2015}, a linear degradation of $G_c$ (and the surface energy) with hydrogen content $\theta$ is assumed:
\begin{equation}
    G_c(\theta) = \left(1-\chi \theta\right)G_c(0).
\end{equation}

\noindent Here, $G_c (0)$ is the critical energy release rate in an inert environment and $\chi$ is the hydrogen damage coefficient, to be calibrated with DFT calculations. For example, based on \citep{Jiang2004a}, $\chi$ equals 0.89 in iron and 0.67 in aluminum. Finally, we make use of the Langmuir-McLean isotherm to compute the hydrogen surface coverage $\theta$ from the bulk hydrogen concentration $C$ as:
\begin{equation}\label{eq:HydrogenOccupancy}
    \theta = \dfrac{C}{C+\exp{(-\Delta g_b^0/RT})},
\end{equation}

\noindent with $\Delta g_b^0$ denoting the difference in Gibbs free energy between the decohering surface and the surrounding material. Assuming that fracture in the presence of hydrogen is intergranular, a value of 30 kJ/mol is assigned to $\Delta g_b^0$ based on the spectrum of experimental data available for the trapping energy at grain boundaries \citep{Serebrinsky2004}. 

\section{Numerical implementation}
\label{Sec:Numerical}

The main features of the finite element framework are introduced in this Section, with further details being provided in \ref{Sec:AppendixFEMDetails}. First, a history field and a strain energy split are defined to prevent damage reversibility and damage under compressive loading (Section \ref{subsec:Split}). Secondly, in Section \ref{subsec:DisFEM} we address the discretisation of the mixed finite element problem and formulate the residuals. In Section \ref{subsec:Visco} we introduce the \textit{ad hoc} viscoplastic law adopted. Finally, the new penalty-based chemical boundary conditions are presented in Section \ref{subsec:ChemBC}. The implementation is conducted within an Abaqus user-element (UEL) subroutine, with the pre-processing of the input files carried out using Abaqus2Matlab \citep{AES2017}.

\subsection{Addressing damage in compression, irreversibility and crack interpenetration}
\label{subsec:Split}

First, a decomposition of the elastic strain energy density is adopted to prevent damage due to compressive stresses. We choose to follow the spherical/deviatoric split first introduced by \citet{Amor2009}. Thus, in a solid with Lame's first parameter $\lambda$, the elastic strain energy density can be decomposed as $\psi^e=\psi_+^e+\psi_{-}^e$, with
\begin{equation}
    \begin{split}
        \psi_+^e &= \frac{1}{2}\left(\lambda +\frac{2}{3}Q \right) \left\langle \tm{tr }\bm{\varepsilon}^e\right\rangle_+^2 + Q \, |{\bm{\varepsilon}^e}^{'}|^2\\
        \psi_{-}^e&=\frac{1}{2}\left(\lambda +\frac{2}{3}Q \right) \left\langle \tm{tr }\bm{\varepsilon}^e\right\rangle_-^2,
    \end{split}
\end{equation}

\noindent and only $\psi_+^e$ contributing to damage. Here, $\left\langle \right\rangle$ denote the Macaulay brackets. The strain energy decomposition is implemented by means of a hybrid approach, following \citet{Ambati2015}. Damage irreversibility, $\phi_{t+\Delta t} \geq \phi_t$, is ensured by introducing a history variable field $H$ \citep{Miehe2010a}. Thus, for a total time $\tau$, the history variable at time $t$ corresponds to the maximum value of $\psi_+^e$, i.e.:
\begin{equation}
    H = \max_{t \in[0,\tau]}\psi^e_+( t). 
\end{equation}

In addition, crack interpenetration is precluded by adding the following constraint \citep{Ambati2015}
\begin{equation}
    \phi=0 \hspace{1cm} \tm{if } \psi_+^e < \psi_{-}^e.
\end{equation}

\subsection{Finite element discretisation}
\label{subsec:DisFEM}

Making use of Voigt notation, the nodal variables for the displacement field, $\mathbf{\hat{u}}$, the plastic strains $\hat{\bm{\varepsilon}}^p$, the phase field, $\hat{\phi}$, and the hydrogen concentration, $\hat{C}$ are interpolated as: 
\begin{equation}\label{eq:Ndiscret}
\mathbf{u} = \sum\limits_{i=1}^m \bm{N}_i^{\bm{u}} \hat{\mathbf{u}}_i, \hspace{1cm} \bm{\varepsilon}^p = \sum\limits_{i=1}^m \bm{N}_i^{\bm{\varepsilon}^p}\hat{\bm{\varepsilon}}^p_i, \hspace{1cm} \phi =  \sum\limits_{i=1}^m N_i \hat{\phi}_i, \hspace{1cm} C =  \sum\limits_{i=1}^m N_i \hat{C}_i.
\end{equation}
\noindent Here, $N_i$ denotes the shape function associated with node $i$, for a total number of nodes $m$. The shape function matrices $\bm{N}^{\bm{u}}_i$ and $\bm{N}_i^{\bm{\varepsilon}^p}$ are given in \ref{Sec:AppendixFEMDetails}. Similarly, the associated gradient quantities can be discretised as:
\begin{equation}\label{eq:Bdiscret}
\bm{\varepsilon} = \sum\limits_{i=1}^m \bm{B}^{\bm{u}}_i \hat{\mathbf{u}}_i, \hspace{0.8cm} \nabla\bm{\varepsilon}^p =  \sum\limits_{i=1}^m \bm{B}^{\bm{\varepsilon}^p}_i \hat{\bm{\varepsilon}}^p_i, \hspace{0.8cm} \nabla\phi =  \sum\limits_{i=1}^m \mathbf{B}_i \hat{\phi}_i, \hspace{0.8cm} \nabla C =  \sum\limits_{i=1}^m \mathbf{B}_i \hat{C}_i,
\end{equation}

\noindent with the \textbf{B}-matrices explicitly given in \ref{Sec:AppendixFEMDetails}.\\

Considering the discretisation (\ref{eq:Ndiscret})-(\ref{eq:Bdiscret}), we derive the residuals for each primal kinematic variable from (\ref{eq:ExternalVirtualWork}) and (\ref{eq:InternalVirtualWork}) as:\\

\noindent \textbullet \- Linear momentum
\begin{equation}
\mathbf{R}_i^\mathbf{u} = \int_\Omega \left\{\left[\left(1-\phi\right)^2+ k\right]\left(\bm{B}^\mathbf{u}_i\right)^T \bm{\sigma}_0 \right\} \, \text{d}V - \int_{\partial\Omega} \left[ \left(\bm{N}^\mathbf{u}_i\right)^T\mathbf{T} \right] \, \text{d}S,
\end{equation}

\noindent where $\bm{\sigma}_0$ is the undamaged stress tensor and $k$ is a small positive parameter introduced to circumvent the complete degradation of the energy. We choose $k=1 \times 10^{-7}$ to ensure that the  algebraic conditioning number remains well-posed for fully-broken states.\\

\noindent \textbullet \- Microplasticity 
\begin{equation}
\mathbf{R}_i^{\bm{\varepsilon}^p} = \int_{\Omega}\left[\left(\bm{N}^{\bm{\varepsilon}^p}_i\right)^T\left(\bm{q} - \bm{\sigma}\right) + \left(\bm{B}^{\bm{\varepsilon}^p}_i\right)^T\bm{\tau}\right] \, \text{d}V-\int_{\partial \Omega} \left[ \left(\bm{N}^{\bm{\varepsilon}^p}_i\right)^T \bm{t} \right] \, \text{d}S.
\end{equation}
\vspace{0.1cm}

\noindent \textbullet \- Phase field
\begin{equation}
R_i^\phi = \int_\Omega \left\{ -2\left(1-\phi\right)N_iH + G_c (C) \left[\frac{\phi}{\ell}N_i + \ell \,  \left( \mathbf{B}_i \right)^T \nabla\phi\right]\right\}dV.
\end{equation}

\noindent where $H$ is the history field variable introduced in Section \ref{subsec:Split}.\\ 

\noindent \textbullet \- Hydrogen transport
\begin{equation}\label{eq:ResidualH}
R_i^C = \int_{\Omega}\left[N_i\left(\dfrac{1}{D}\dfrac{dC}{dt}\right) + \mathbf{B}_i^T\nabla C - \mathbf{B}_i^T\left(\dfrac{\overline{V}_H C}{RT}\nabla\sigma_H\right)\right] \, \text{d}V + \dfrac{1}{D}\int_{\partial\Omega_\rho} N_i\rho \, \text{d}S.
\end{equation} 

The consistent tangent stiffness matrices $\bm{K}$, required to complete the finite element implementation, are obtained by considering the constitutive relations and differentiating the residuals with respect to the incremental nodal variables; details are given in \ref{Sec:AppendixFEMDetails}. For each element, the linearised weakly-coupled system reads:
\begin{equation}\label{eq:CompleteSystem}
\begin{bmatrix}
\bm{K}^{\mathbf{u},\mathbf{u}} & \bm{K}^{\mathbf{u},\bm{\varepsilon}^p} & 0 & 0 \\
\bm{K}^{\bm{\varepsilon}^p,\mathbf{u}} & \bm{K}^{\bm{\varepsilon}^p,\bm{\varepsilon}^p} & 0 & 0 \\
0 & 0 & \bm{K}^{\phi,\phi} & 0 \\
0 & 0 &0 & \bm{K}^{C,C}
\end{bmatrix}\begin{bmatrix}
\mathbf{u} \\ \bm{\varepsilon}^p \\ \phi \\ C
\end{bmatrix} + \begin{bmatrix}
0 & 0 & 0 & 0\\ 0 & 0 & 0 & 0\\ 0 & 0 & 0 & 0\\ 0 & 0 & 0 & \bm{M}
\end{bmatrix} \begin{bmatrix}
0 \\ 0 \\ 0 \\ \dot{C} \end{bmatrix}= \begin{bmatrix}
\mathbf{R}^\mathbf{u} \\ \mathbf{R}^{\bm{\varepsilon}^p} \\ R^\phi \\ R^C
\end{bmatrix},
\end{equation}

\noindent where $\bm{M}=\partial R^C_i/ \partial\dot{C}$ is the concentration capacity matrix. The global finite element system is solved, in an implicit time integration framework, by means of a staggered approach, following the work by \citet{Miehe2010a}. A time sensitivity study is conducted in all computations.

\subsection{Viscoplastic law}
\label{subsec:Visco}

We circumvent the need to track active plastic regions (see, \citealp{Nielsen2013}) by employing a viscoplastic function that is particularized to the rate-independent limit. Thereby, the effective stress $\Sigma$ is defined as:
\begin{equation}
\Sigma= \sigma_F\left(E^p\right)V\left(\dot{E}^p\right),
\end{equation}

\noindent where $\sigma_F$ is the current flow stress and $V(\dot{E}^p)$ is the viscoplastic function. We assume isotropic power-law hardening such that, for a strain hardening coefficient $N$, the flow rule reads:
\begin{equation}
\sigma_F = \sigma_y\left(1+\dfrac{E^p}{\varepsilon_y}\right)^N. \label{Powerlaw}
\end{equation}

\noindent Here, $\sigma_y$ denotes the initial yield stress, and accordingly the yield strain is given by $\varepsilon_y=E/\sigma_y$. For a reference strain rate $\dot{\varepsilon}_0$ and strain rate sensitivity exponent $m$, the viscoplastic function reads:
\begin{equation}\label{eq:viscoplastic1}
    V \left(\dot{E}^p\right) = \left( \frac{\dot{E}^p}{\dot{\varepsilon}_0} \right)^m.
\end{equation}

\noindent As discussed in Section \ref{Sec:SmallScaleYielding}, the values of $\dot{\varepsilon}_0$ and $m$ are appropriately chosen so as to reproduce rate-independent behavior.\\

The widely used viscoplastic hardening rule (\ref{eq:viscoplastic1}) is implemented by adopting the viscoplastic split recently proposed by \citet{IJP2020}, which builds on the previous work by \citet{Panteghini2016}. The aim is to bound the magnitude of $\partial \Sigma / \partial \Delta E^p$ when $\dot{E}^p \to 0$, preventing ill-conditioning. The definition in (\ref{eq:viscoplastic1}) is approximated by:
\begin{equation}
    V \left(\dot{E}^p\right) =   \begin{cases}
    \frac{\dot{E}^p}{\varpi \dot{\varepsilon}_0}     & \quad \text{if } \dot{E}^pm/\dot{E}^p_* \leq 1\\
    \left(\frac{\dot{E}^p-\frac{1-m}{m}\dot{E}^p_*}{\dot{\varepsilon}_0}\right)^m & \quad \text{if } \dot{E}^pm/\dot{E}^p_* > 1
  \end{cases}
\end{equation}.

\noindent Here, $\varpi$ is a small positive constant ($\varpi<<1$) and $\dot{E}^p_*$ is a threshold quantity that is defined to ensure a smooth transition between states:
\begin{equation}
    \dot{E}^p_* = \dot{\varepsilon}_0 \left( \frac{1}{\varpi m} \right)^{1/(m-1)}.
\end{equation}

\noindent More details are given in \citep{IJP2020}. 

\subsection{Chemical conditions on a moving boundary}
\label{subsec:ChemBC}

\begin{figure}
    \centering
    \includegraphics{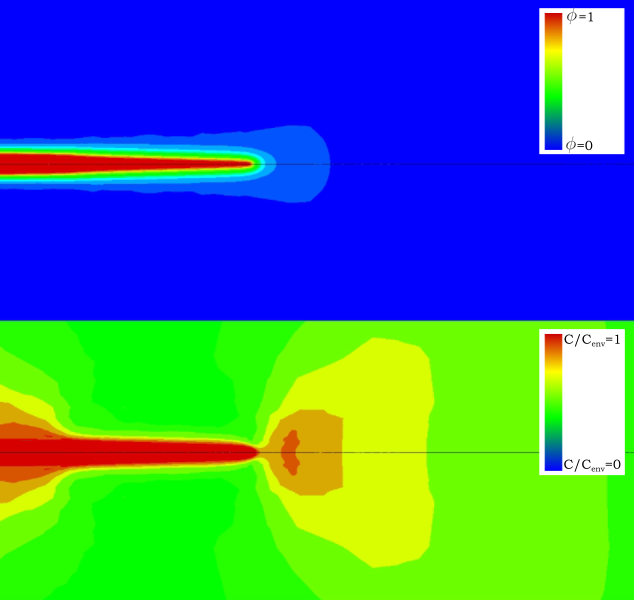}
    \caption{\emph{Moving} chemical boundary conditions with a propagating crack, contours of phase field damage (top) and hydrogen concentration (bottom). Details of the boundary value problem are given in Section \ref{Sec:Rcurves}.}
    \label{fig:Penalty}
\end{figure}

Consider a solid that is continuously exposed to a hydrogenous environment. Fracture modeling requires capturing how the environment-solid boundary advances with crack growth. The newly formed crack surface is promptly exposed to the environment, hydrogen gas or an aqueous electrolyte. The use of a phase field framework facilitates tracking crack advance and, accordingly, prescribing suitable chemical conditions on the moving boundary. Specifically, we make use of a penalty-based approach \citep{Renard2020}, and include a penalty term $P$ multiplying the first term of the hydrogen concentration residual (\ref{eq:ResidualH}), with the associated stiffness matrix being modified accordingly. The penalty term is defined in terms of a large positive penalty coefficient $k_p\approx 1\times 10^5$ as: 
\begin{equation}\label{eq:penalty}
    P = k_p (C-C_{env})\langle \phi- 0.5 \rangle_{+} ,
\end{equation}

\noindent such that the hydrogen concentration at the crack surface approaches the environmental hydrogen concentration $C_{env}$ as $\phi \to 1$. Representative contours of crack advance, as defined by $\phi=1$, and hydrogen concentration are given in Fig. \ref{fig:Penalty}. The role of the environment in providing a continuous source of hydrogen is captured.   

\section{Results}
\label{Sec:Results}

The capabilities of the model will be demonstrated by addressing representative case studies. First, in Section \ref{Sec:Stationary} the role of the plastic length scales on stationary crack tip fields is investigated. Secondly, physical insight will be gained by exploring the relation between crack growth resistance and fracture process parameters in a wide variety of scenarios. Crack growth resistance curves are computed to explore the sensitivity of the model to (i) the plastic length scale parameters, (ii) the fracture length scale parameter, (iii) the hydrogen concentration, and (iv) the rate of loading, see Section \ref{Sec:Rcurves}. In addition, the steady state fracture toughness is estimated as a function of the strength, showing that the model can naturally capture the ductile-to-brittle transition experienced in the presence of hydrogen. Finally, the capabilities of the model in quantitatively capturing experimental results are showcased in Section \ref{Sec:CaseStudy0} by comparing with crack initiation measurements, $K_{th}$, under a wide range of environments (applied potentials, $E_p$). 

\subsection{Mode I fracture of an elastic-plastic solid in the presence of hydrogen}
\label{Sec:SmallScaleYielding}

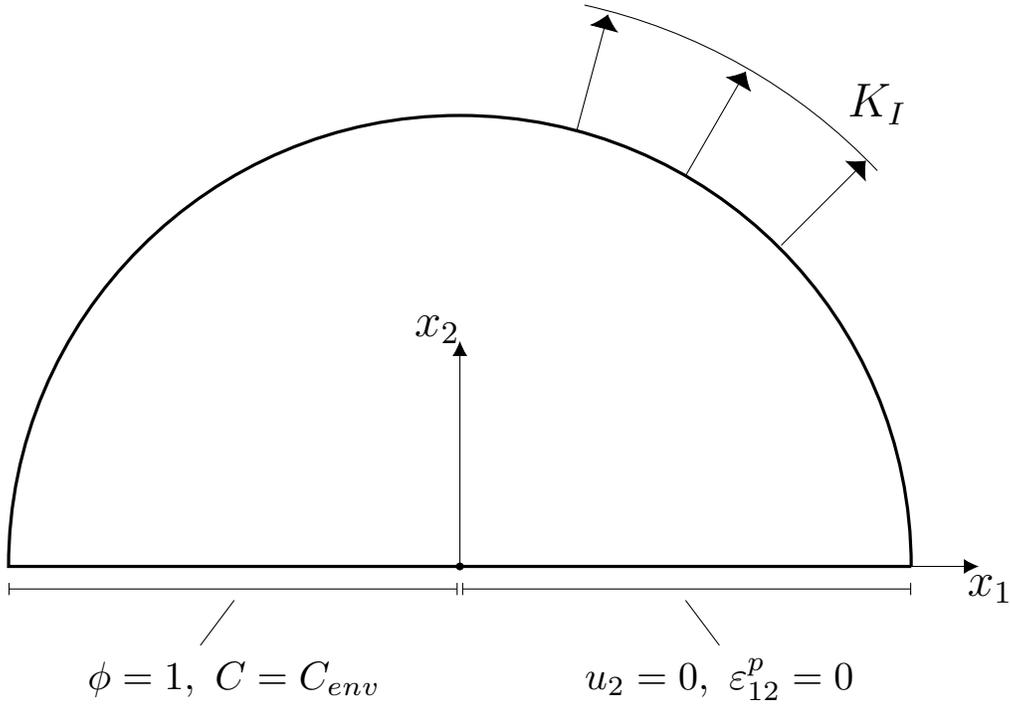
\begin{figure}[H]
    \centering
\begin{tikzpicture}[scale=1.5, every node/.style={transform shape}]
\draw [very thick] (0,0) --(4,0);
\draw [very thick] (4,0) arc [radius=4, start angle=0, end angle=180] -- (0,0);
\draw [-{Latex[length=2mm,width=2mm]}] (0,0) -- (4.6,0);
\draw [-{Latex[length=2mm,width=2mm]}] (0,0) -- (0,2);
\node at (4.7,-0.2) {$x_1$};
\node at (-0.2,2.1) {$x_2$};
\draw [fill=black] (0,0) circle [radius=0.03];
\draw [-{Latex[length=2mm,width=4mm]}] (2.8484,2.8484) -- (3.6,3.6);
\draw [-{Latex[length=2mm,width=4mm]}] (2,3.4641) -- (2.535,4.3907);
\draw [-{Latex[length=2mm,width=4mm]}] (1.0352,3.8637) -- (1.3122,4.8972);
\draw (3.6994,3.5106) arc [radius=5.1, start angle=43.5, end angle=77.5];
\node at (3.7,4.1) {$K_I$};
\draw[|-|] (-4,-0.2) -- (-0.02,-0.2);
\draw[|-|] (4,-0.2) -- (0.02,-0.2);
\draw (-2.3,-0.7) -- (-2,-0.3);
\draw (2.3,-0.7) -- (2,-0.3);
\node at (-2.,-1) {\footnotesize $\phi=1,\tm{ } C=C_{env}$};
\node at (2.3,-1) {\footnotesize $u_2=0,\tm{ } \varepsilon^p_{12}=0$};
\end{tikzpicture}
    \caption{Sketch of the boundary layer formulation and the associated mechanical, chemical and damage boundary conditions.}
    \label{fig:ModelSketches}
\end{figure}

We assume that small scale yielding conditions prevail and make use of a boundary layer formulation to prescribe a remote $K_I$ field, see Fig. \ref{fig:ModelSketches}. Consider a crack with its tip at the origin of the coordinate system and with the crack plane along the negative axis of the Cartesian reference frame $(x_1,x_2)$. The elastic response of the solid is characterised by the Young's modulus $E$ and Poisson's ratio $\nu$. Then, an outer $K_I$ field is imposed by prescribing nodal displacements on the outer periphery of the mesh as
\begin{equation}
u_i = \frac{K_I}{E} r^{1/2} f_i \left( \theta, \nu \right),
\end{equation}
\noindent where the subscript index $i$ equals $x_1$ or $x_2$, and the functions $f_i \left( \theta, \nu \right)$ are given, in terms of polar coordinates $(r, \theta)$ centred at the crack tip, by
\begin{equation}
f_{1} = \frac{1+\nu}{\sqrt{2 \pi}} \left(3 - 4 \nu - \cos \theta \right) \, \cos \left(\frac{\theta}{2} \right)
\end{equation}
\noindent and
\begin{equation}
f_{2} = \frac{1+\nu}{\sqrt{2 \pi}} \left(3 - 4 \nu - \cos \theta \right) \, \sin \left(\frac{\theta}{2} \right).
\end{equation}

Upon exploiting reflective symmetry about the crack plane, only half of the finite element model is analysed. We note in passing that satisfying reflective symmetry requires careful consideration of the higher order boundary conditions. Consider (\ref{eq:balance_BC})b; for a crack lying on the $x_2$ axis, micro-free boundary conditions $\bm{t}=\bm{0}$ imply, along the $x_2=0$ plane:
\begin{equation}\label{eq:symmetryBCs}
    \tau_{222}=\tau_{112}=\tau_{122}=0.
\end{equation}

\noindent Symmetry dictates that $\varepsilon^p_{22,2}=\varepsilon^p_{11,2}=0$ along the extended crack path, such that the Neumann boundary conditions $t_{11}=t_{22}=0$ are appropriate. However, $\varepsilon^p_{12}$ is an odd function in $x_2$ which requires prescribing instead $\varepsilon^p_{12}=0$ at the symmetry plane, as the conventional Neumann boundary condition $\sigma_{12}=0$ at $x_2=0$ does not imply $\varepsilon^p_{12}=0$ for non-zero values of $L_D$ and $L_E$.\\

The boundary layer formulation will be employed to shed light into the role of plastic strain gradients on stationary crack tip fields, and it is subsequently used to characterise crack growth resistance in embrittled elastic-plastic solids.

\subsubsection{Stationary crack tip fields}
\label{Sec:Stationary}

Consider a semi-infinite, stationary crack in an elastic-plastic solid subjected to a remote elastic $K_I$ and in the absence of hydrogen. A representative value for the plastic zone size $R_p$ can be estimated from Irwin's approximation as:
\begin{equation}
R_p=\frac{1}{3 \pi} \left( \frac{K_I}{\sigma_y}\right)^2
\end{equation}

The viscoplastic parameters $\dot{\varepsilon}_0$ and $m$ are chosen to ensure that we are close the rate-independent limit, as confirmed by comparison with conventional rate-independent plasticity predictions for $L_E=L_D=0$.\\ 

Of interest here is the behavior of the opening tensile stresses, $\sigma_{22}$, and the hydrostatic stress, $\sigma_H$, relevant to both fracture and hydrogen transport. In the rate independent limit, any stress quantity ahead of the crack is a function of the following non-dimensional parameters:
\begin{equation}
    \frac{\sigma}{\sigma_y} = F \left( \frac{x_1}{R_p}, \, \frac{L_E}{R_p}, \, \frac{L_D}{R_p}, \, N, \, \nu, \, \frac{E}{\sigma_y} \right).
\end{equation}

\begin{figure}[H]
    \centering
    \begin{subfigure}{0.6\textwidth}
    \includegraphics[width=\textwidth]{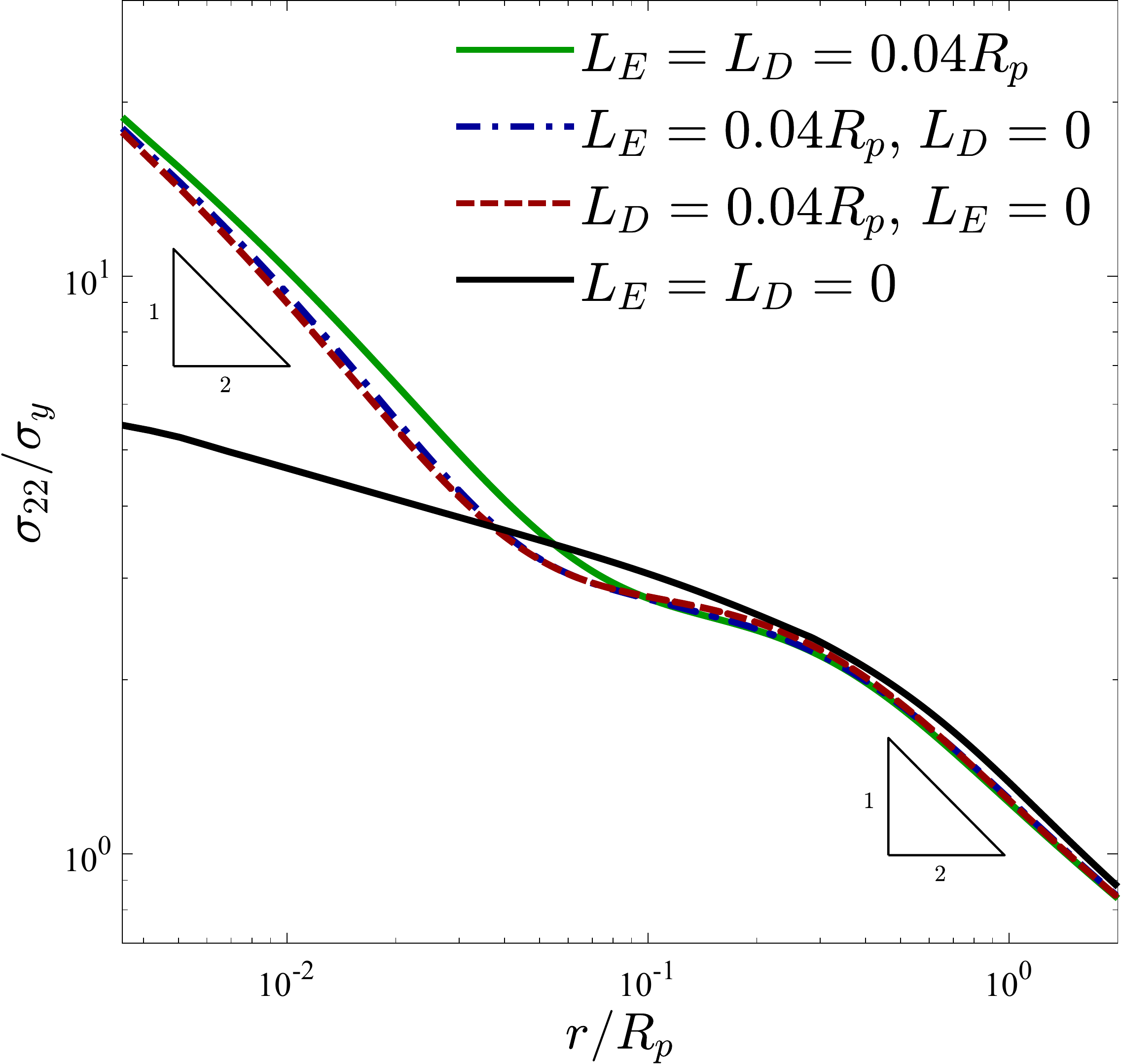}
    \caption{ }
    \label{fig:stationaryCracka}
    \end{subfigure}
        \begin{subfigure}{0.6\textwidth}
 \includegraphics[width=\textwidth]{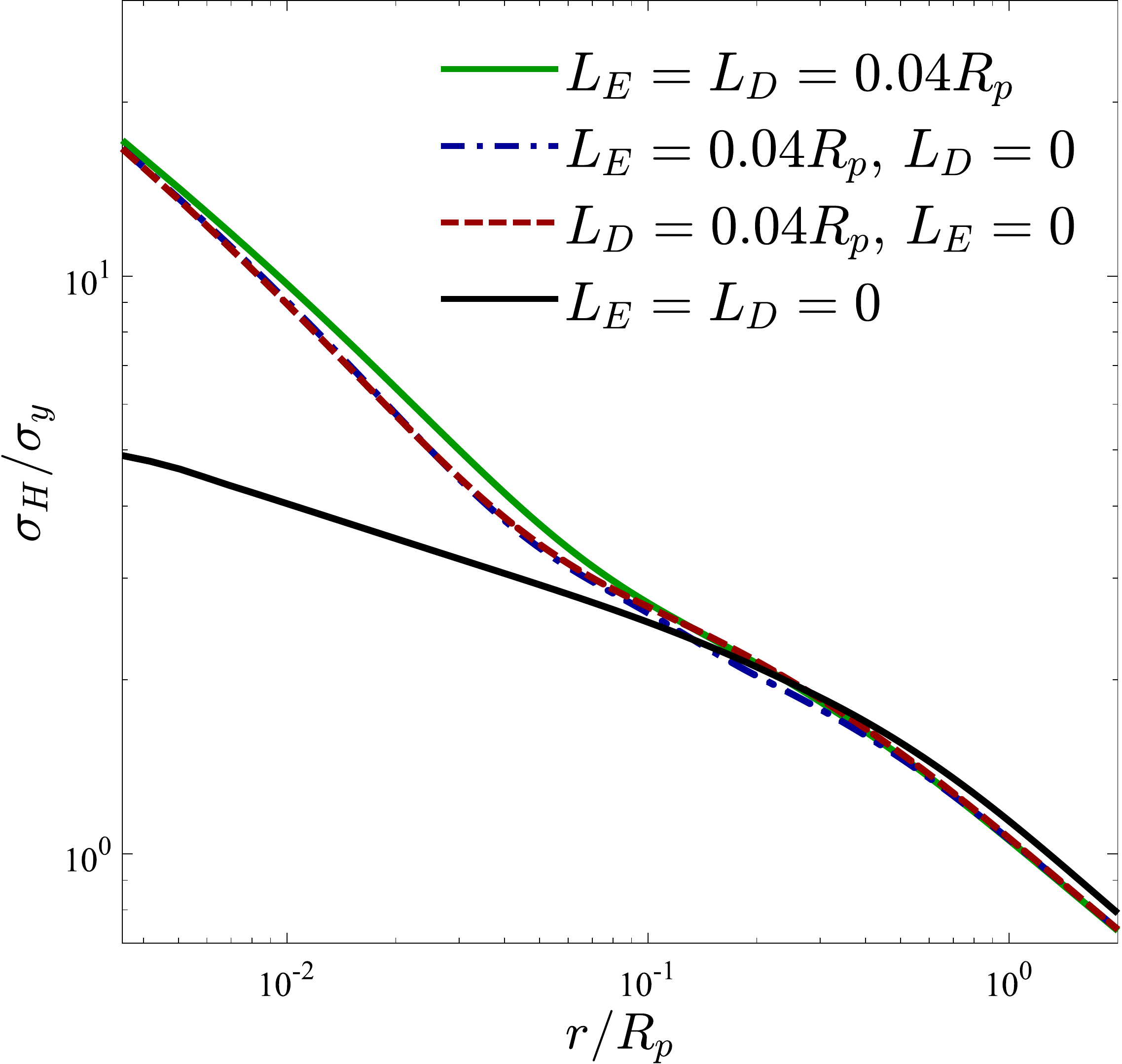}
     \caption{ }
    \label{fig:stationaryCrackb}
    \end{subfigure}
    \caption{Crack tip fields ahead of a stationary crack in the absence of hydrogen: (a) opening tensile stress distribution, and (b) hydrostatic stress distribution. Material properties: $\sigma_y/E=0.003$, $\nu=0.3$ and $N=0.2$.}
    \label{fig:StationaryCrack}
\end{figure}

The numerical results obtained for $\sigma_{22}/\sigma_y$ and $\sigma_H/\sigma_y$ ahead of the crack tip are shown in a log-log scale in Figs. \ref{fig:StationaryCrack}a and \ref{fig:StationaryCrack}b, respectively. Results are obtained for a solid with $\sigma_y/E=0.003$, $\nu=0.3$ and strain hardening exponent $N=0.2$. Regarding the plastic length scales, different combinations are considered: (i) conventional plasticity, $L_E=L_D=0$; (ii) purely energetic hardening $L_E=0.04R_p$ (with $L_D=0$); (iii) purely dissipative strengthening $L_D=0.04R_p$ (with $L_E=0$); and (iv) combined dissipative and energetic strengthening, $L_E=L_D=0.04R_p$. Consider first the case of the opening tensile stress distribution, Fig. \ref{fig:StationaryCrack}a. Away from the crack tip, predictions agree, independently of the value of $L_D$ and $L_E$. However, strain gradient plasticity predictions lead to much higher stresses than those obtained with conventional plasticity as we approach the crack tip. In fact, for all cases when $L_D$ or $L_E$ are non-zero, the finite element results reveal the existence of an inner $K_I$-field, where the stress field recovers the linear elastic $r^{-1/2}$ singularity. The existence of this \emph{elastic core} has been recently justified analytically by \citet{EJMAS2019} and it is reminiscent of a dislocation free crack tip zone, as introduced by \citet{Suo1993}. Note that the inner $K_I$-field is present for any non-zero choice of $L_E$ and $L_D$, with the purely energetic result predicting slightly higher stresses than the purely dissipative case but with differences being minimal. For the nearly-proportional loading conditions of the stationary crack tip problem, differences are due to the different weighting of energetic and dissipative higher order contributions, see (\ref{eq:qDandTauD}b) and (\ref{eq:tauE}). 

The hydrostatic stress $\sigma_H$ distribution is shown in \ref{fig:StationaryCrack}b and reveals the same qualitative trends. First, the stress distribution predicted by strain gradient plasticity agrees with the conventional plasticity result far from the crack tip, but predictions start to differ when $r$ is on the order of the relevant plastic length scale. Also, similar results are obtained when strain gradient contributions are purely energetic ($L_E > 0, \, L_D = 0$) and purely dissipative ($L_D > 0, \, L_E =0 $). In both cases, a substantial stress elevation is attained close to the crack tip, where the hydrostatic stress level is roughly four times larger than the conventional plasticity prediction. This could have important implications for modeling hydrogen transport, given the exponential dependence of the hydrogen concentration on the hydrostatic stress. For example, for a given hydrogen concentration at the boundary $C_{env}$, the hydrogen concentration $C$ at steady state reads \citep{Liu1970}:
\begin{equation}
    C=C_{env} \exp \left( \frac{\bar{V}_H \sigma_H}{R T} \right)
\end{equation}

\noindent The prediction of large hydrogen concentrations within a few microns of the crack tip surface is consistent with experimental measurements, see \citet{IJHE2016} and \citet{Gerberich2012}.

\subsubsection{Crack growth resistance}
\label{Sec:Rcurves}

Consider now the case of a growing crack, as dictated by the phase field. Insight will be first gained on the role of the plastic length scales, and the effect of hydrogen will be subsequently taken into consideration. A fracture process zone length $R_0$ can be defined, as done by \citet{Tvergaard1992} in the context of cohesive zone models, as
\begin{equation}\label{eq:R0}
    R_0 = \dfrac{1}{3\pi\left(1-\nu^2\right)}\dfrac{EG_c}{\sigma_y^2},
\end{equation}

In addition, a cohesive bonding strength can be defined to frame the ductile versus brittle dichotomy. In the context of phase field models, a critical stress $\hat{\sigma}$ can be defined from the homogeneous solution to (\ref{eq:PhaseFieldStrongForm}) in a one dimensional setting. As shown by, for example, \citet{Borden2012} and \citet{CMAME2018}, this cohesive strength can be expressed as a function of Young's modulus $E$, the fracture energy $G_c$ and the phase field length scale $\ell$ as:
\begin{equation}\label{eq:criticalstress}
    \hat{\sigma} = \dfrac{9}{16}\sqrt{\dfrac{EG_c}{3\ell}}.
\end{equation}

Thus, $\ell$ is a material parameter that determines the magnitude of the critical stress. Crack initiation is based on a purely energetic criterion, $G=G_c$, but crack growth resistance will be affected by the material strength, as determined through $\ell$. Also, given (\ref{eq:criticalstress}), the material strength $\hat{\sigma}$ will decrease with increasing hydrogen content \textit{via} its relation with $G_c (C)$. Using (\ref{eq:criticalstress}) one can establish an analogy with cohesive zone models, where the traction-separation law is characterised by its shape, the value of the fracture energy $G_c$ and the cohesive strength $\hat{\sigma}$. However, we emphasize that, in general, equation (\ref{eq:criticalstress}) constitutes an approximation. Accordingly, we choose to favour using $\ell/R_0$ as a relevant non-dimensional group, which is inversely related to $\hat{\sigma}/\sigma_y$, the common choice in cohesive zone analyses, as:
\begin{equation}
\frac{R_0}{\ell} = \frac{256}{81 \pi \left( 1-\nu^2 \right)} \left( \frac{\hat{\sigma}}{\sigma_y} \right)^2    
\end{equation}

Thus, in the absence of hydrogen and in the rate-independent limit, dimensional analysis implies that the crack growth resistance depends on the following non-dimensional groups:
\begin{equation}
\dfrac{K_I}{K_0} = F\left(\dfrac{\Delta a}{R_0},\tm{ }\dfrac{\ell}{R_0},\tm{ }N,\tm{ }\dfrac{E}{\sigma_y},\tm{ }\dfrac{L_p}{R_0},\tm{ } \nu \right),
\end{equation}

\noindent where $L_p$ is the reference plastic length scale: $L_E=L_D=L_p$, $\Delta a$ is the crack extension, and $K_0$ is the reference stress intensity factor at which cracking initiates. Under plane strain conditions, the remote load at which cracking initiates is given by:
\begin{equation}
    K_0 = \left( \frac{E G_c}{1-\nu^2} \right)^{1/2}
\end{equation}

Results are computed using the boundary layer formulation shown in Fig. \ref{fig:ModelSketches}. The crack is introduced by prescribing the phase field parameter, $\phi=1$. We assume plane strain conditions and the domain is discretised using a total of approximately 36000 quadratic quadrilateral elements with reduced integration. The characteristic element length along the crack propagation path is at least 6 times smaller than the phase field length scale $\ell$, so as to resolve the fracture process zone and ensure mesh insensitive results \citep{CMAME2018}. Material properties are given by $\sigma_y/E=0.003$, $\nu=0.3$ and $N=0.2$, unless otherwise stated. The viscoplastic parameters are chosen to model the rate-independent limit. Specifically, we define the following dimensionless constant:
\begin{equation*}
    c_r = \dfrac{\dot{K}_I \varepsilon_y}{K_0\dot{\varepsilon}_0},
\end{equation*}

\noindent and make suitable choices for $c_r$ and $m$. The combination $c_r=0.24$ and $m=0.025$ reproduces the rate-independent limit, as confirmed by comparing with the results obtained with rate-independent J2 plasticity (for $L_p=0$) and with the viscoplastic function by \citet{Panteghini2016}.\\

\begin{figure}[H]
    \centering
    \includegraphics[width=\textwidth]{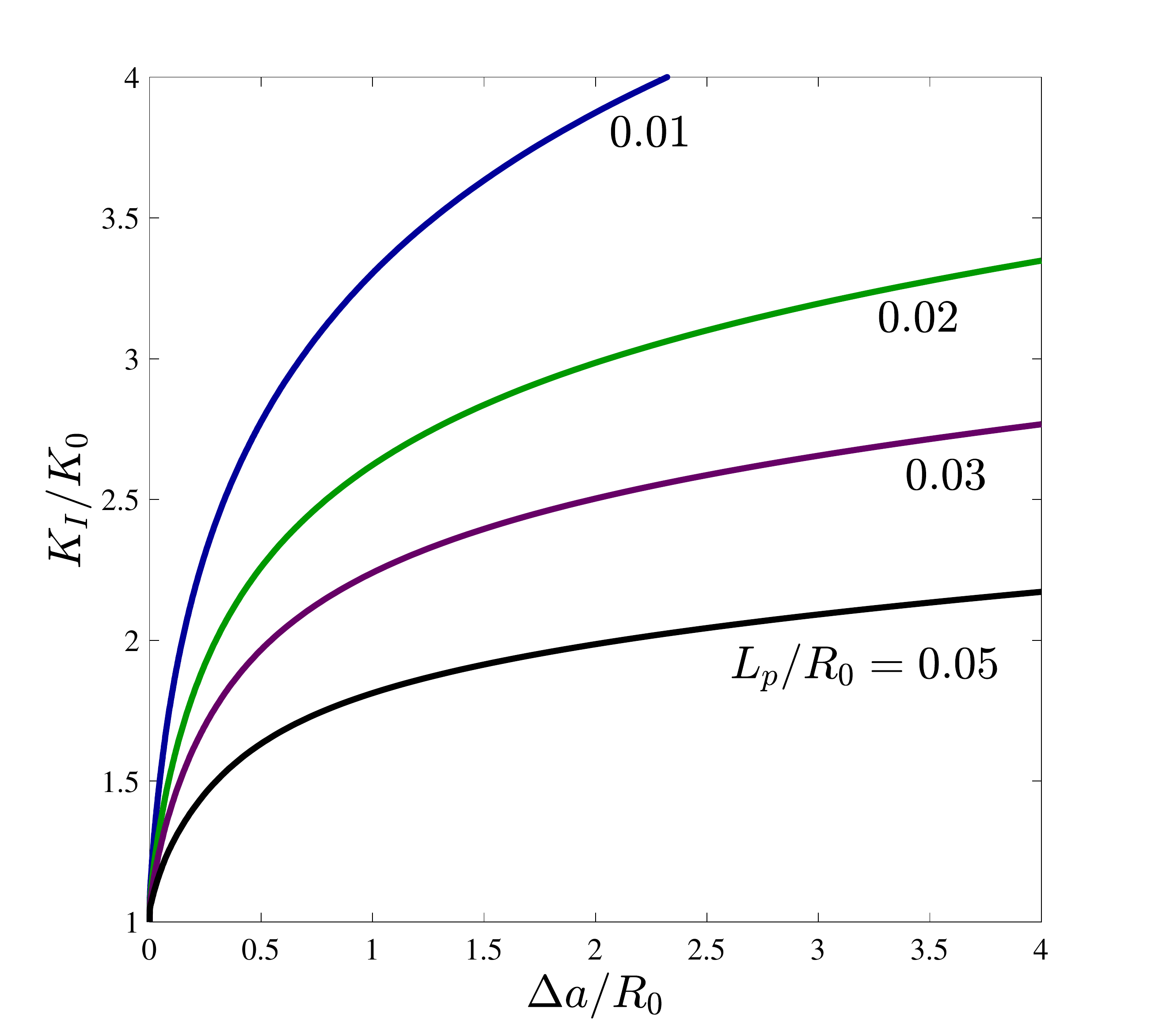}
    \caption{Influence of the plastic length scale on crack growth resistance. Material properties: $\sigma_y/E=0.003$, $\nu=0.3$, $N=0.2$, and $\ell/R_0=1/35$ $\left(\hat{\sigma}/{\sigma_y}\approx 5.6\right)$.}
    \label{fig:PlasticLengthScale}
\end{figure}

First, the influence of the plastic length scale on the fracture resistance is assessed. As shown in Fig. \ref{fig:PlasticLengthScale}, crack growth resistance curves (R-curves) are computed for selected values of $L_p/R_0$. In agreement with expectations, larger values of $L_p/R_0$ magnify gradient effects, elevating crack tip stresses and reducing the fracture resistance. The precise magnitude of $L_p/R_0$ depends mainly on the potential of the material to strengthen or harden in the presence of plastic strain gradients, as given by $L_p$, and on the work of fracture, as given by $G_c$ - see (\ref{eq:R0}). The fracture energy $G_c$ can vary from a few J/m$^2$, as in fracture processes governed by atomic decohesion, to hundreds of kJ/m$^2$, as in ductile damage. Consequently, dislocation hardening effects have a higher influence in brittle cracking, where the work of separation and the fracture process zone are small.

\begin{figure}[H]
    \centering
    \includegraphics[width=\textwidth]{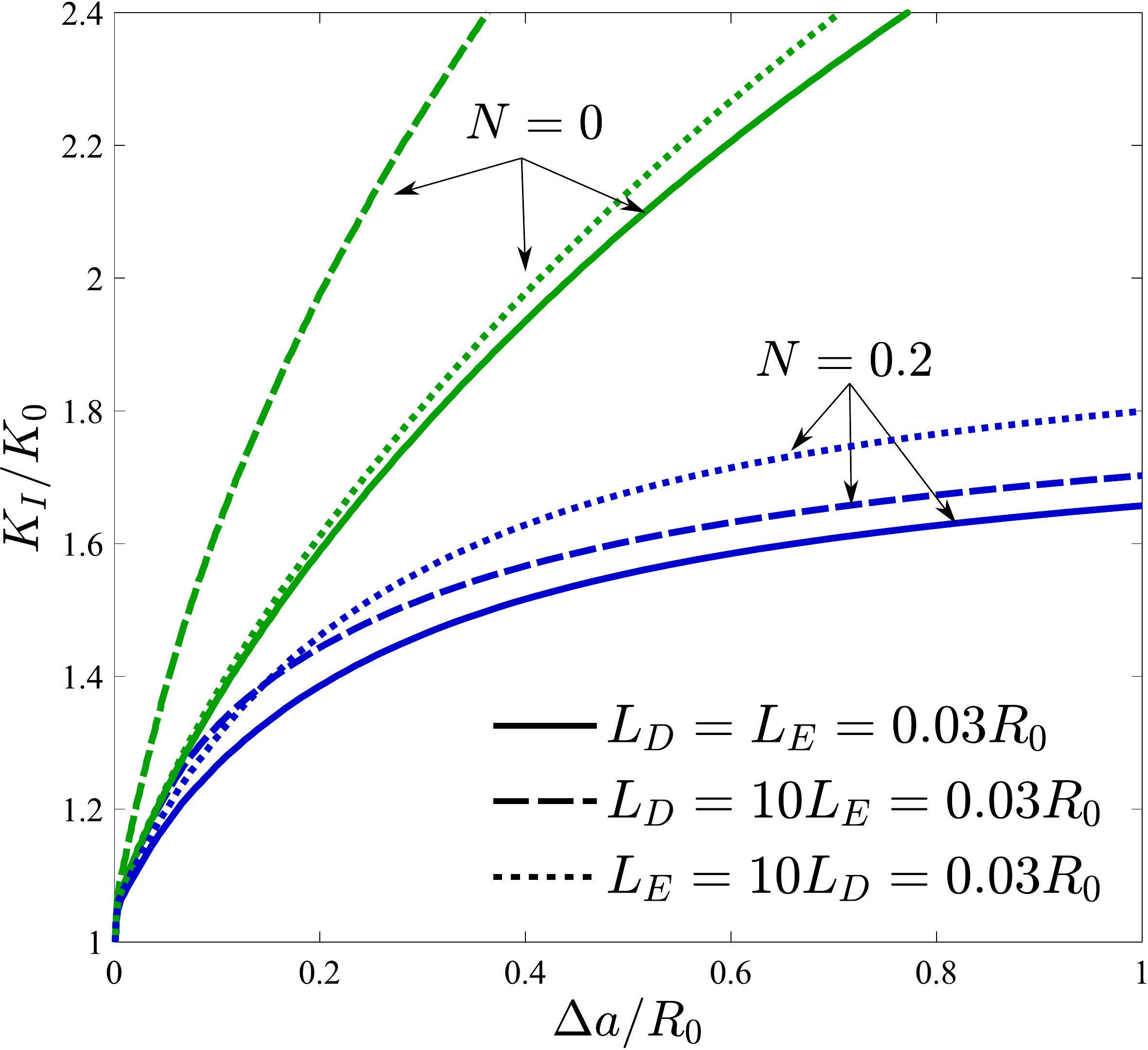}
    \caption{Individual influence of the energetic and dissipative plastic length scales on crack growth resistance. Material properties: $\sigma_y/E=0.003$, $\nu=0.3$, and $\ell/R_0=1/15$ $\left(\hat{\sigma}/{\sigma_y}\approx 3.7\right)$. Green curves correspond to $N=0$ while blue curves denote the $N=0.2$ case.}
    \label{fig:IndividualLengthScales}
\end{figure}

We aim at elucidating the contributions of the individual energetic and dissipative plastic length scales to the reduction in fracture resistance with increasing $L_p/R_0$ shown in Fig. \ref{fig:PlasticLengthScale}. As non-proportional straining becomes relevant with crack advance, the expectation is to observe larger differences than those reported in Fig. \ref{fig:StationaryCrack} for the stationary crack. Crack growth resistance curves are shown in Fig. \ref{fig:IndividualLengthScales} for three cases: (i) $L_E=10L_D=0.03R_0$, (ii) $L_D=10L_E=0.03R_0$, and (iii) $L_D=L_E=0.03R_0$ (the reference case). Results are given for two choices of the strain hardening exponent: $N=0.2$ and $N=0$ (perfectly plastic behavior). As it would be expected, the less steep R-curves for both $N=0$ and $N=0.2$ cases are given by the combined energetic and dissipative strengthening results, $L_D=L_E=0.03R_0$. Interestingly, dissipative effects appear to dominate the response for $N=0.2$, while energetic contributions are more significant in the $N=0$ case. Differences between energetic-dominated ($L_E >> L_D$) and dissipative-dominated ($L_D >> L_E$) predictions are due to the constitutive definitions of their associated higher order stresses, $\bm{\tau}^E$ and $\bm{\tau}^D$, see (\ref{eq:qDandTauD}b) and (\ref{eq:tauE}). While $\bm{\tau}^D$ is related to the plastic strain gradients through a power-law expression, $\bm{\tau}^E$ is related to $\nabla \bm{\varepsilon}^p$ by a linear relation. As evident from Fig. \ref{fig:IndividualLengthScales}, fracture takes place at smaller loads and thereby at smaller plastic strains for $N=0.2$, a domain where $\bm{\tau}^D$ will dominate. Conversely, for a fixed $\ell/R_0$ ($\hat{\sigma}/\sigma_y$), much larger strains and plastic dissipation take place in the case where $N=0$. In addition, one should note that differences may also arise due to the kinematic nature of the energetic contribution, which resembles a back-stress \citep{Legarth2010}. As shown recently by \citet{JAM2018} and \citet{EFM2019} in the context of conventional plasticity, kinematic hardening increases plastic dissipation and fracture resistance, relative to isotropic hardening.

\begin{figure}[H]
    \centering
    \includegraphics[width=1.1\textwidth]{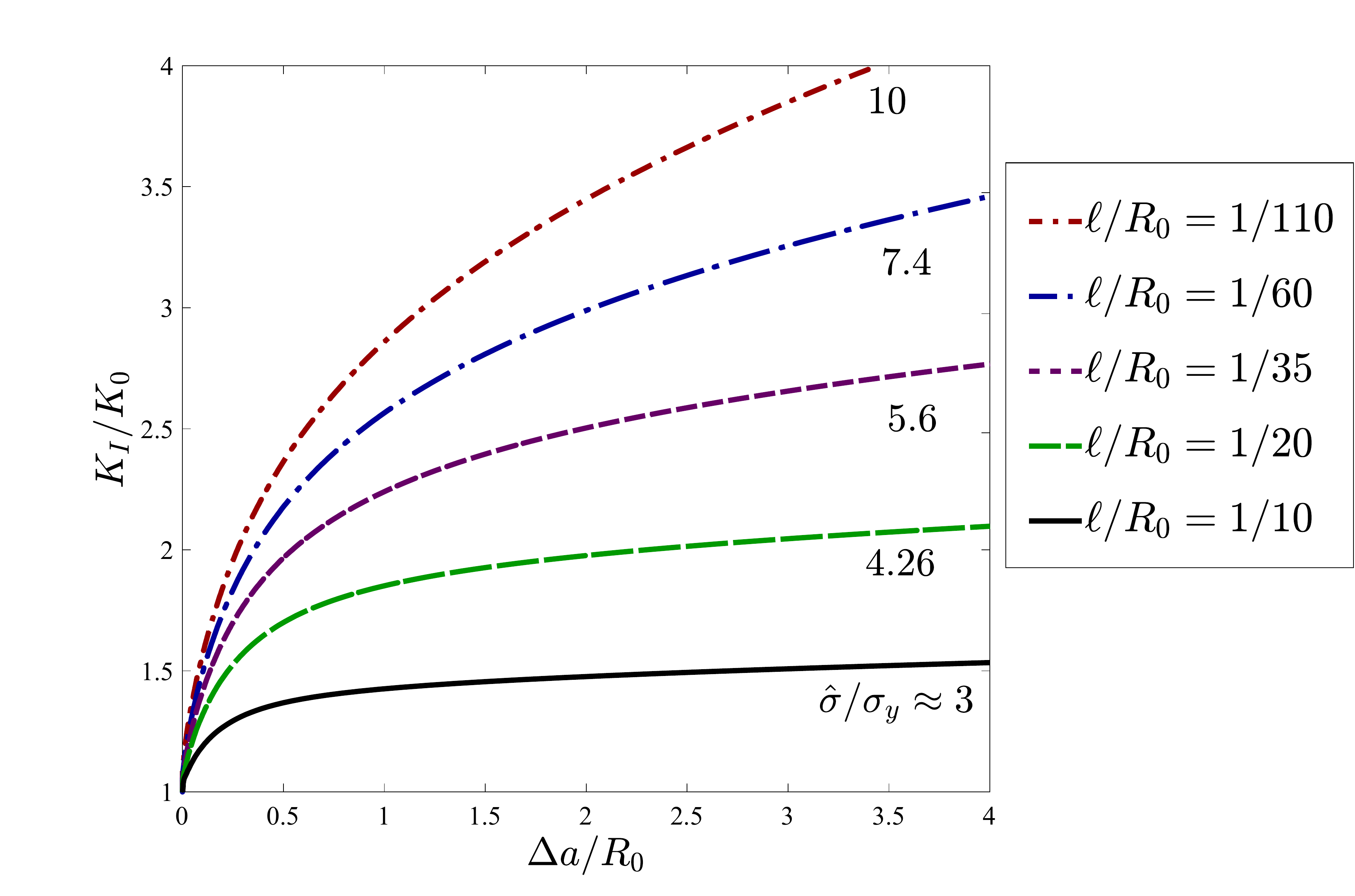}
    \caption{Influence of the strength $\hat{\sigma}/\sigma_y$ ($\ell/R_0$) on the crack growth resistance. Material properties: $\sigma_y/E=0.003$, $\nu=0.3$, $N=0.2$ and $L_p/R_0=0.03$. }
    \label{fig:CriticalStress}
\end{figure}

We proceed to vary the $\ell/R_0$ ratio to explore the sensitivity of the fracture resistance to the critical stress (see (\ref{eq:criticalstress})). As shown in Fig. \ref{fig:CriticalStress}, augmenting $\ell/R_0$ (or $\hat{\sigma}/\sigma_y$) increases the steepness of the R-curve. This qualitative trend agrees with the results obtained by \citet{Tvergaard1992} using cohesive zone models. Since atomic decohesion requires attaining $\hat{\sigma}/\sigma_y$ values on the order of 10 or larger (as opposed to ductile fracture, $\hat{\sigma}/\sigma_y \approx 4$), the magnitude of the phase field length scale can be tailored to capture a specific cracking mechanism. The results shown in Fig. \ref{fig:CriticalStress} span a wide range of scenarios, with $\ell<L_p$ and $L_p<\ell$. \citet{Miehe2016b} have chosen the phase field length scale to be smaller than the plastic length scale, on the grounds of a regularised crack zone lying inside the plastic zone. However, we emphasize that the magnitude of $\ell$ does not correspond to the width of the crack smearing function, and that $L_p$ is a (constant) material property that does not correspond to the size of the plastic zone. More importantly, Fig. \ref{fig:CriticalStress} shows that, if $L_p/R_0$ is sufficiently large, fracture can be attained at critical stresses on the order of the theoretical lattice strength $\hat{\sigma}=10\sigma_y$. Thus, atomic decohesion in the presence of plasticity, as observed in numerous material systems (\citealp{Elssner1994}; \citealp{Bagchi1996}; \citealp{Korn2002}), can be rationalised in the context of strain gradient plasticity. This is unlike conventional plasticity, where crack tip stresses are only 3-5 larger than the initial yield stress and, consequently, fracture does not occur if the cohesive strength $\hat{\sigma}$ is on the order of the theoretical lattice strength ($\approx 10 \sigma_y$) or the grain boundary strength ($\approx 9 \sigma_y$), see (\citealp{Tvergaard1992}; \citealp{Duda2015}). In the absence of hydrogen, the magnitude of $L_p/R_0$ has to be increased up to 0.03 to predict quasi-cleavage, i.e. fracture with $\hat{\sigma}=10 \sigma_y$. Note that $L_p/R_0$ values in the 0.001-0.01 range are expected for ductile steels; $L_p$ is a material property that can be measured from micro-scale experiments, with $L_p \approx 1-10$ $\mu$m for most metals, and $R_0$ is on the order of 1 mm or more for void controlled fracture processes, where a work of fracture of tens of kJ/m$^2$ is at least required \citep{Wei1997}. Such values of $L_p/R_0$ will be insufficient to trigger cleavage fracture in the absence of hydrogen, as crack growth is governed by other mechanisms, like void nucleation, growth and coalescence. However, hydrogen significantly reduces the work of fracture $G_c$, entailing an increase in the magnitude of $L_p/R_0$ that can trigger the ductile to brittle transition observed in the experiments.


\begin{figure}[H]
    \centering
    \includegraphics[width=\textwidth]{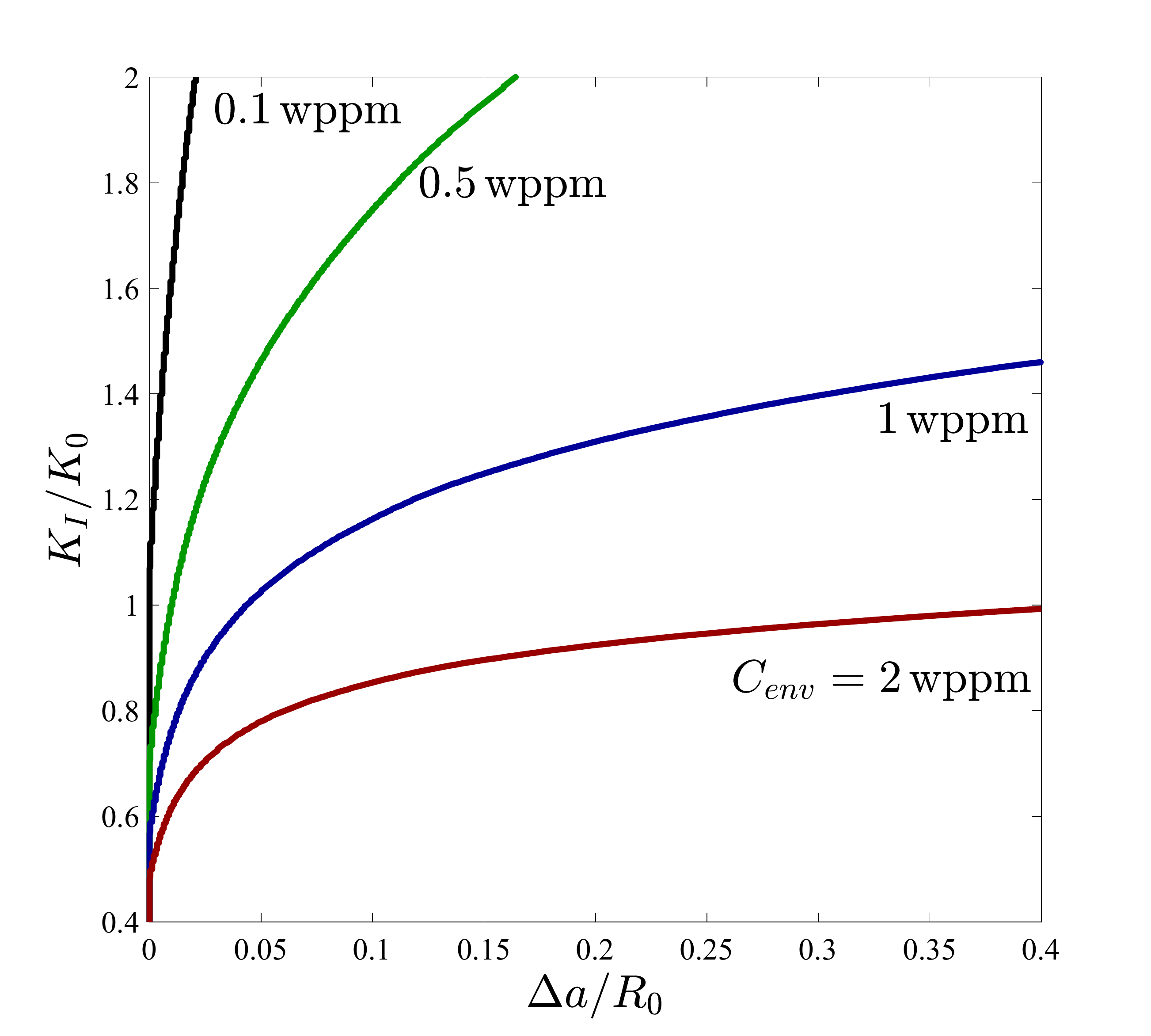}
    \caption{Influence of the environmental hydrogen concentration on the fracture resistance. Material properties: $\sigma_y/E=0.003$, $\nu=0.3$, $N=0.2$, $\ell/R_0=1/110$ $(\hat{\sigma}/\sigma_y=10)$, $L_p/R_0=0.001$, $D=0.0127$ mm$^2$/s and $\chi=0.89$. Loading rate $\dot{K}_I/K_0=4 \times 10^{-7} \, \tm{s}^{-1}$.}
    \label{fig:FigHconcentration}
\end{figure}

Consider now the influence of hydrogen. Can we predict cleavage failure ($\hat{\sigma}=10 \sigma_y$) for $L_p/R_0$ values that are realistic for ductile steels? We proceed to compute crack growth resistance curves with $\ell/R_0=1/110$, $L_p/R_0=0.001$ and selected values of the environmental hydrogen concentration $C_{env}=0.1$, $0.5$, $1$, and $2$ wppm. $C_{env}$ is prescribed at the crack surfaces and the specimen is not initially pre-charged, such that hydrogen charging and mechanical loading start at the same time. A value for the diffusion coefficient typical of iron-based materials is assumed, $D=0.0127$ mm$^2$/s, following \citet{Sofronis1989}. The remote loading is prescribed at a rate of $\dot{K}_I/K_0=4 \times 10^{-7} \, \tm{s}^{-1}$ and the hydrogen damage coefficient is chosen to be $\chi=0.89$, based on the atomistic calculations by \citet{Jiang2004a} for hydrogen in Fe. The predictions obtained are shown in Fig. \ref{fig:FigHconcentration}; the model appropriately captures the observed trend of a decreasing fracture resistance with increasing hydrogen concentration. We emphasize that no hydrogen pre-charging is considered and the results are reported relative to the initial (inert) values of $K_0$ and $R_0$. Thus, cracking initiates below $K_I/K_0=1$ in all cases, with the magnitude of $K_I/K_0$ at crack initiation decreasing with increasing $C_{env}$. By incorporating the role of hydrogen, fracture is predicted assuming cleavage cracking, $\hat{\sigma}=10 \sigma_y$, in an otherwise ductile material, $L_p/R_0=0.001$. The implications of this finding will be discussed below by computing the relation between the steady state fracture toughness $K_{SS}$ and the critical cohesive strength $\hat{\sigma}$.

\begin{figure}[H]
    \centering
    \includegraphics[width=0.75\textwidth]{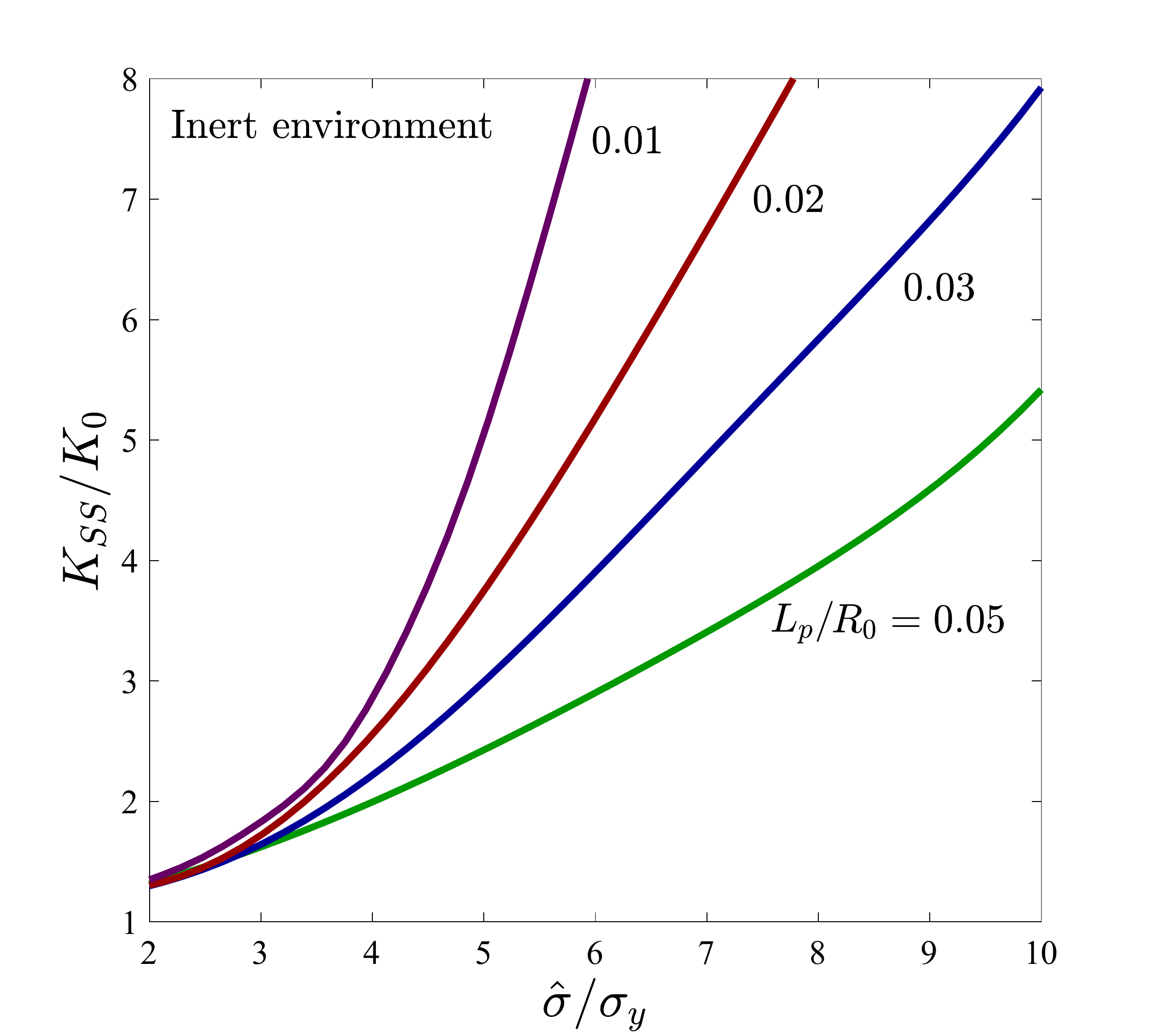}
        \includegraphics[width=0.75\textwidth]{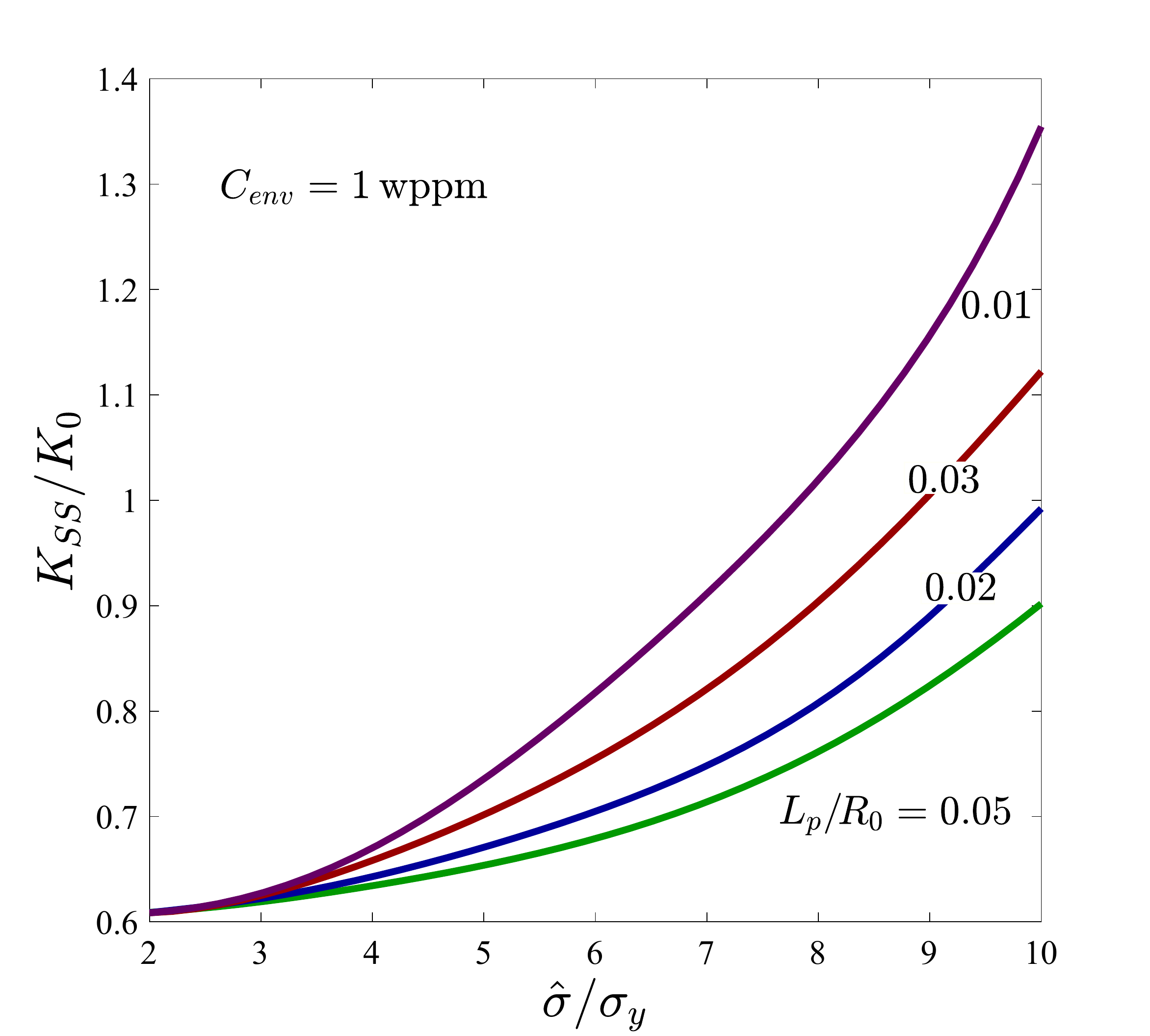}
    \caption{Steady state fracture toughness $K_{SS}/K_0$ as a function of the cohesive strength $\hat{\sigma}/\sigma_y$ for selected values of $L_p/R_0$: (a) inert environment, and (b) hydrogenous environment, with $C_{env}=1$ wppm. Material properties: $\sigma_y/E=0.003$, $\nu=0.3$, $N=0.2$, $D=0.0127$  mm$^2$/s and $\chi=0.89$. Loading rate $\dot{K}_I/K_0=4 \times 10^{-7} \, \tm{s}^{-1}$. }
    \label{fig:SteadyState}
\end{figure}

Steady state curves are shown in Fig. \ref{fig:SteadyState} for both an inert and a hydrogenous environment ($C_{env}=1$ wppm). The steady state fracture toughness is estimated from the R-curves, with $K_{SS}$ being the limiting value attained by $K_I$ as the crack approaches steady state. The magnitude of $K_{SS}/K_0$ is computed for a wide range of strengths ($\ell/R_0$) and selected values of $L_p/R_0$. Consider first the results in the absence of hydrogen, Fig. \ref{fig:SteadyState}a. The model predicts ductile fracture for values of $L_p/R_0$ below 0.03 ($\hat{\sigma}/\sigma_y<10$). As discussed above, this is consistent with the magnitude of $R_0$ in ductile metals, as given by the work of fracture. On the other hand, brittle fracture is predicted for high values of $L_p/R_0$, as it is the case in metal-ceramic interfaces \citep{Odowd1992}, ferritic steels at low temperatures (\citealp{Qian2011}; \citealp{EJMaS2019b}) and other material systems (see, e.g., \citealp{Wang1991}) where the fracture energy is on the order of 1 kJ/m$^2$ or lower. However, the response changes drastically when hydrogen is taken into consideration, see Fig. \ref{fig:SteadyState}b. Even for the case of ductile metals, $L_p/R_0 \approx 0.01$, the steady state curve intersects the brittle fracture threshold, $\hat{\sigma}/\sigma_y=10$. Thus, the transition from ductile to brittle fracture observed in the experiments is captured.

\begin{figure}[H]
    \centering
    \includegraphics[width=\textwidth]{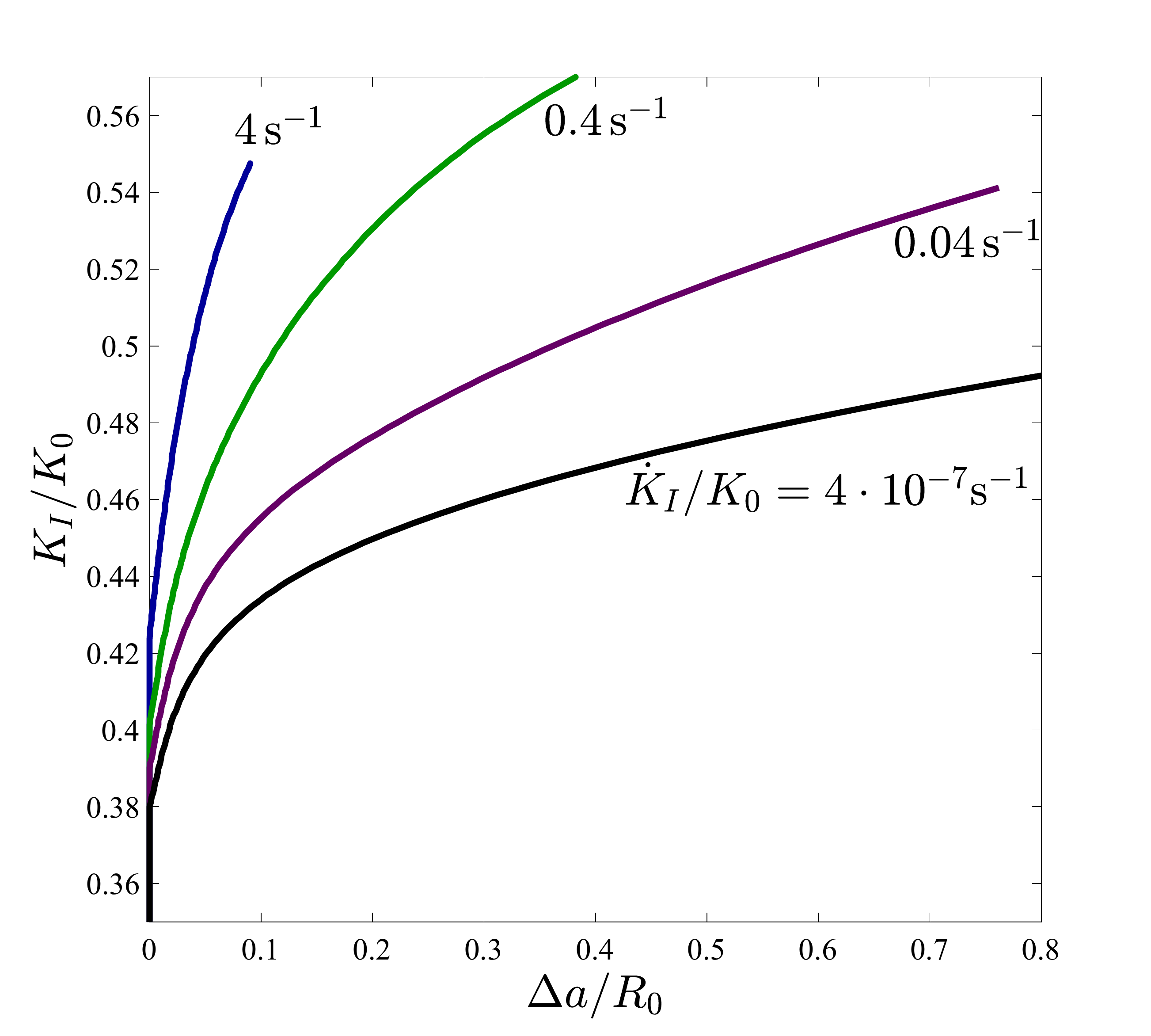}
    \caption{Influence of the loading rate $\dot{K}_I/K_0$ on the fracture resistance in a sample pre-charged uniformly with a hydrogen concentration of 1 wppm. Material properties: $\sigma_y/E=0.003$, $\nu=0.3$, $N=0.2$, $\ell/R_0=1/60$, $L_p/R_0=0.01$, $D=0.0127$ mm$^2$/s and $\chi=0.89$. }
    \label{fig:HydrogenTime}
\end{figure}

In addition, we show that the model is also capable of predicting internal hydrogen assisted cracking and the sensitivity of the R-curve to the loading rate. Thus, a uniform hydrogen pre-charging of 1 wppm is assumed, and crack growth resistance curves are computed for selected values of the loading rate $\dot{K}_I/K_0$. The results, shown in Fig. \ref{fig:HydrogenTime}, exhibit the expected trends: slower loading rates emphasize embrittlement and lead to less steep R-curves. As we increase the loading rate there is less time for the hydrogen to diffuse to the fracture process zone, where $\sigma_H$ is large, and the hydrogen-induced degradation of the local fracture energy is less severe (relative to slower loading rates).  

The crack growth resistance results presented show that the model can rationalise and capture the transition to brittle fracture due to hydrogen, as well as reproducing the main experimental trends (sensitivity to loading rate and hydrogen concentration). 

\subsection{Comparison with experiments: predicting the onset of cracking in ultra-high strength steel (AerMet100)}
\label{Sec:CaseStudy0}

We proceed to compare model predictions with experimental measurements of stress intensity thresholds $K_{th}$ for crack initiation. This analysis is inspired by the encouraging agreement with experiments on ultra-high alloys obtained by \citet{AM2016}. In their work, strain gradient plasticity analyses of crack tip fields and electrochemical assessment of hydrogen solubility were integrated into \citet{Gerberich2012} dislocation-based model. Cracking thresholds $K_{th}$ and stage II crack growth rates $da/dt_{II}$ predictions showed a very good agreement with experiments conducted over a wide range of applied potentials on a nickel superalloy, Monel K-500, and on an ultra-high strength steel, AerMet100. Here, we seek to demonstrate the same capability for our proposed model which explicitly models both cracking and hydrogen transport. Attention is limited to the case of the modern ultra-high strength steel AerMet100 and the estimation of the threshold stress intensity factor $K_{th}$.\\

As detailed in \citep{Lee2007,Pioszak2017}, pre-cracked fracture mechanics specimens were subjected to slowly increasing mode I loading, while submerged in an aqueous solution of 0.6 M NaCl. Loading is feedback controlled such that after an initial loading to 6 MPa$\sqrt{\tm{m}}$, the loading rate is held constant at $\dot{K}= 6.8\cdot10^{-4}$ MPa$\sqrt{\tm{m}}$/s. A wide range of environments are considered, with the applied potential ranging from -1.1 to -0.5 V\textsubscript{SCE}. The measured material properties of AerMet100 are given in Table \ref{tab:AerMetParams} \citep{Lee2007}.
\begin{table}[H]
    \centering
    \begin{tabular}{c c c c c}
        $E$ [GPa] & $\nu$ [-] & $\sigma_y$ [MPa] & $N$ [-] & $D$ [cm\textsuperscript{2}/s] \\\hline\hline
       194 & 0.3 & 1725 & 0.077 & $1 \times 10^{-9}$
    \end{tabular}
    \caption{Material parameters for AerMet100.}
    \label{tab:AerMetParams}
\end{table}

The reference plastic length scale is chosen to be equal to $L_p=5$ $\mu$m, an intermediate value within the range of experimentally measured length scales reported in the literature \citep{IJP2020}. The phase field length scale $\ell$ is chosen appropriately such that the material strength corresponds to the cleavage strength $\hat{\sigma} / \sigma_y=10$. Regarding the fracture and hydrogen damage properties, the choice of $G_c$ (or $K_0$) will establish the maximum value of $K_{th}$ that can be reached. Since our model and choice of material strength aim at reproducing quasi-cleavage, we choose $K_0$ to approximately match the maximum value of $K_{th}$ that is attained in the experiments without observing ductile fracture features. According to \citet{Pioszak2017}, when soluble hydrogen is below 0.8 wppm, corresponding to the potential range $E_p = -0.770$ V\textsubscript{SCE} to $-0.667$ V\textsubscript{SCE}, part of the fracture process is controlled by void coalescence. Hence, we assume $K_0=30$ MPa$\sqrt{m}$. The hydrogen damage coefficient $\chi$ is calibrated to provide a best fit to the experiments and our choice is subsequently discussed.\\

Small scale yielding conditions are assumed, and a boundary layer formulation is employed, as illustrated in Fig. \ref{fig:ModelSketches}. The remote $K$-field is applied at a constant rate of $\dot{K}=7.0\cdot10^{-4}$ MPa$\sqrt{m}$/s. The rate-independent limit for the viscoplastic law is attained using the measures described in section \ref{Sec:Rcurves}. The diffusible hydrogen concentration associated with each value of the applied potential is obtained from the analysis by \citet{Kehler2008}. For an applied potential $E_p$ below -0.75 V\textsubscript{SCE}, both the upper and lower bounds of the crack tip soluble hydrogen are given by: 
\begin{equation}
    \label{eq:CenvLowerBound}
    C_{env}\tm{(wppm)} = 19.125E_p^3 + 78.568E_p^2 + 80.026E_p + 24.560 \,\tm{(V\textsubscript{SCE})}
\end{equation}
For potentials above -0.75V\textsubscript{SCE}, the upper bound solution is applied, which is given by: 
\begin{align}
  \label{eq:CenvUpperBound}
   C_{env}\tm{(wppm)} = {-739.24}E_p^5-3121.1E_p^4&-5147.1E_p^3-4099.2E_p^2\\&-1563.8E_p-225.77 \, \tm{(V\textsubscript{SCE})}\nonumber
\end{align}

\begin{figure}[H]
    \centering
    \includegraphics[width=0.75\textwidth]{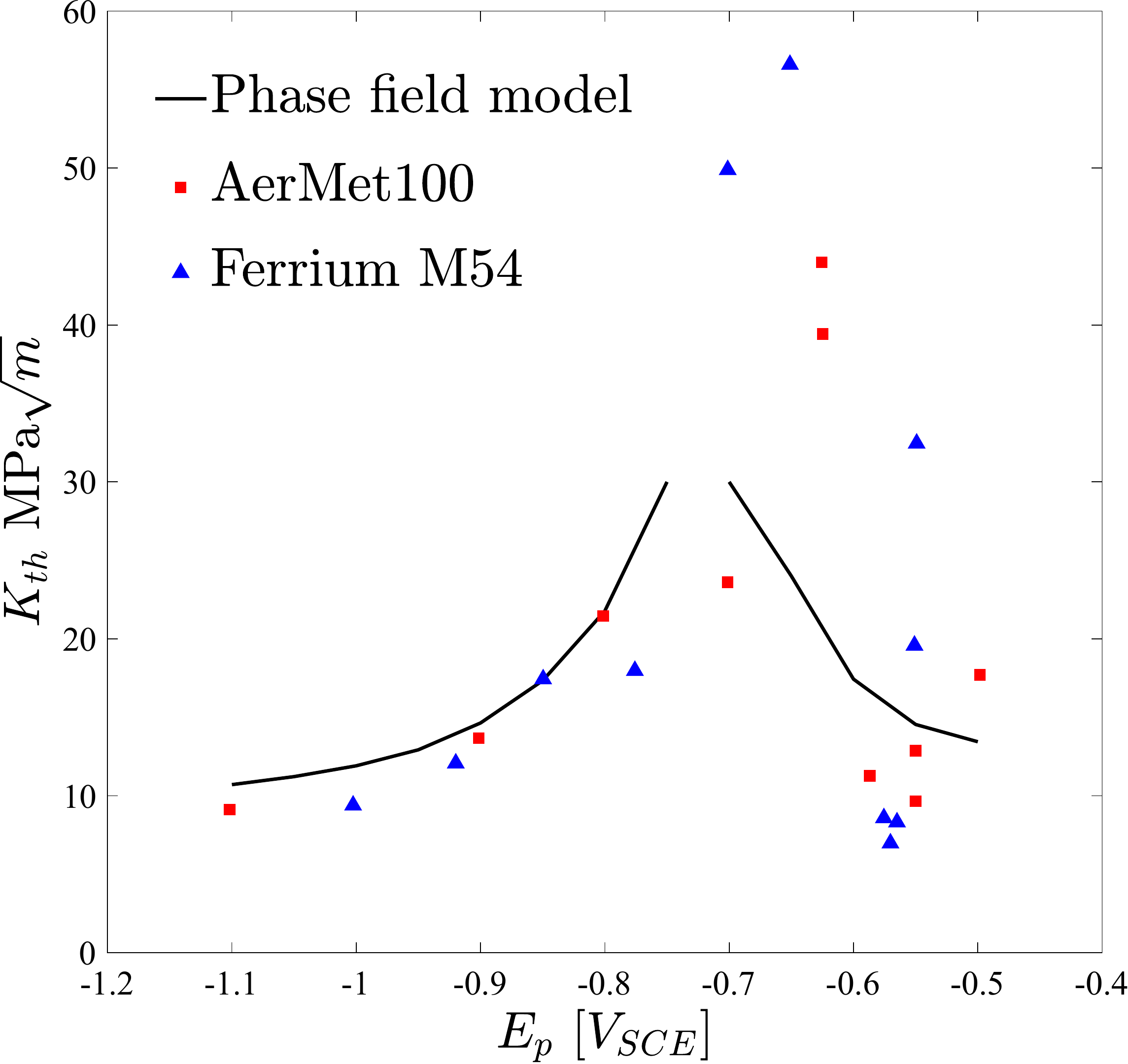}
    \caption{Stress intensity threshold $K_{th}$ predictions as a function of the applied potential. The experimental results obtained in AerMet100 and Ferrium M54 by Gangloff and co-workers \citep{Lee2007,AM2016,Pioszak2017} are shown for comparison. Ductile fracture features are observed in the range $E_p = -0.770$ V\textsubscript{SCE} to $-0.667$ V\textsubscript{SCE}.}
    \label{fig:AerMet}
\end{figure}

The crack initiation threshold $K_{th}$ predictions of the present model are shown in  Fig. \ref{fig:AerMet}, along with experimental results for AerMet100 and Ferrium M54, a similar alloy \citep{Pioszak2017,AM2016}. A very good agreement with experiments is observed over the range of potentials where fracture is reported as quasi-brittle. The best fit is given for a choice of the hydrogen damage coefficient equal to $\chi=0.97$. This value is above the first principles estimate for iron, $\chi=0.89$, see \citep{Jiang2004a,CMAME2018}. However, the choice of the hydrogen damage coefficient that best fits the experimental results is sensitive to the choice of trap binding energy; $\Delta g_b^0=30$ kJ/mol in Eq. (\ref{eq:HydrogenOccupancy}). Traps with higher binding energies than 30 kJ/mol are likely to be present in AerMet100 and participate in the fracture process \citep{Li2004}.

\section{Conclusions}
\label{Sec:Conclusions}

We have presented a new coupled deformation-diffusion-fracture theory for modeling hydrogen embrittlement in elastic-plastic solids. The model builds upon a stress-assisted diffusion formulation, driven by chemical potential gradients, and a chemo-mechanical phase field description of fracture. A fracture energy degradation law is defined that establishes a direct and quantitative connection with atomistic calculations of surface energy reduction with hydrogen coverage. More importantly, the model takes into consideration the role of plastic strain gradients and the associated dislocation hardening mechanisms. The aim is to capture the flow strength elevation observed in mechanical tests involving non-homogeneous plastic deformation and in crack tip discrete dislocation dynamics simulations. Both energetic hardening and dissipative strengthening dislocations mechanisms are considered through their associated plastic length scales: $L_E$ and $L_D$, respectively. In addition, the model is also non-local with respect to the damage variable, with a phase field length scale entering the constitutive relations due to dimensional consistency. Starting from the principle of virtual work, chemical-, micro- and macro-force balances are derived, together with a standard free-energy imbalance. The formulation is completed with a set of thermodynamically-consistent constitutive equations for the deformation, diffusion and fracture problems. \\

The coupled model is implemented in a four-field finite element framework, with displacements, plastic strains, hydrogen concentration and phase field parameter being the primary kinematic variables. Suitable \emph{moving} chemical boundary conditions, viscoplastic function and energy split are introduced, and the coupled problem is solved in an implicit time integration scheme. Numerical calculations are conducted to gain fundamental physical insight and showcase the capabilities of the model. First, the role of plastic strain gradients in elevating crack tip stresses and hydrogen concentrations is assessed in the context of stationary cracks. Much higher crack tip stresses and hydrogen concentrations are predicted, relative to conventional plasticity, providing a suitable physical ground for atomic-scale decohesion. Crack growth resistance curves (R-curves) are computed to elucidate the interplay between the different plastic and fracture length scales involved in the formulation. Increasing $L_E$ or $L_D$ relative to the plastic zone size reduces the steepness of the R-curve, as gradient hardening is exacerbated. On the other side, decreasing the magnitude of the phase field length scale is equivalent to augmenting the strength $\hat{\sigma}$, leading to an increased fracture resistance. The steady state fracture toughness is computed as a function of $\hat{\sigma}/\sigma_y$ to gain insight into the underlying fracture mechanism; brittle fracture occurs when the strength is on the order of $10 \sigma_y$ or larger, following lattice and grain boundary strength arguments. Results reveal a high sensitivity to the ratio of the reference plastic length scale, $L_p=L_D=L_E$, to the fracture process zone $R_0$, whose magnitude is mainly governed by the work of fracture $G_c$. In the absence of hydrogen, brittle fracture is only predicted for large values of $L_p/R_0$, characteristic of low fracture energy material systems, such as metal-ceramic interfaces or ferritic steels at low temperatures. However, when hydrogen is taken into consideration, the fracture energy is substantially reduced and brittle fracture is predicted also for ductile metals, where $L_p/R_0$ is initially small. Therefore, a framework is proposed that can rationalise quasi-cleavage in the presence of plasticity and the change from microvoid cracking to brittle fracture observed in ductile steels in the presence of hydrogen. Lastly, the quantitative predictive capabilities of the model have been benchmarked by reproducing mode I fracture experiments on an ultra-high strength steel, AerMet100. The results obtained under a wide range of applied potentials reveal a promising agreement with experiments.

\section{Acknowledgments}
\label{Sec:Acknowledge of funding}

The authors gratefully acknowledge financial support from the Danish Hydrocarbon Research and Technology Centre (DHRTC) under the ''Reliable in-service assessment in aggressive environments'' project, publication number DHRTC-PRP-101. E. Mart\'{\i}nez-Pa\~neda also acknowledges financial support from ExxonMobil, Wolfson College Cambridge (Junior Research Fellowship) and from the Royal Commission for the 1851 Exhibition through their Research Fellowship programme (RF496/2018). C.F. Niordson additionally acknowledges support from the Danish Council for Independent Research through the research project ``Advanced Damage Models with InTrinsic Size Effects'' (Grant no: DFF-7017-00121).


\appendix

\section{Details of numerical implementation}
\label{Sec:AppendixFEMDetails}

Here, we provide explicit expressions for the matrix operators and the stiffness matrix components used in Section \ref{subsec:DisFEM}.

\subsection{Matrix operators}

Assuming plane strain conditions, for a node $i$, the nodal solutions to the deformation problem read:
\begin{equation}
  \hat{\mathbf{u}}_i = [ \hat{u}_1^i , \, \, \hat{u}_2^i ] \,\,\, , \,\,\,\,\,\,\,\,\,\,\,\,\,\,\,\,\,\,\,\, \hat{\bm{\varepsilon}}^p_i = [ {\varepsilon_{11}^p}^i , \, \, {\varepsilon_{22}^p}^i , \, \, {\varepsilon_{12}^p}^i , \, \, {\varepsilon_{13}^p}^i , \, \, {\varepsilon_{23}^p}^i ] .
\end{equation}

\noindent Accordingly, the shape function matrices are given as follows:

\begin{equation}
    \bm{N}_{i}^{\mathbf{u}}=
        \begin{bmatrix}
            N_{i}   &   0 \\[0.3em]
            0   &   N_{i}
        \end{bmatrix}, \hspace{1cm} \bm{N}_i^{\bm{\varepsilon}^p} = \begin{bmatrix}
N_i & 0 & 0 \\ 0 & N_i & 0 \\ -N_i & -N_i & 0 \\ 0& 0 & N_i
\end{bmatrix},
\end{equation} 

\noindent while the gradient quantities are discretised using: 
\begin{equation}
\mathbf{B}_i = \begin{bmatrix}
\frac{\partial N_i}{\partial x} \\[0.1cm] \frac{\partial N_i}{\partial y} 
\end{bmatrix}, \hspace{0.6cm}
\bm{B}_{i}^{\mathbf{u}}=
        \begin{bmatrix}
            \frac{\partial N_i }{\partial x}   &   0   \\[0.3em]
            0   &    \frac{\partial N_i}{\partial y}    \\[0.3em]
            0   &   0    \\[0.3em]
            \frac{\partial N_i}{ \partial y}  &  \frac{ \partial N_i}{ \partial x} 
        \end{bmatrix}, \hspace{0.6cm} 
 \bm{B}_i^{\bm{\varepsilon}^p} = \begin{bmatrix}
\frac{\partial N_i}{\partial x} & 0 & 0 \\[0.1cm] 
\frac{\partial N_i}{\partial y} & 0 & 0 \\[0.1cm] 
0 & \frac{\partial N_i}{\partial x} & 0 \\[0.1cm]
0 & \frac{\partial N_i}{\partial y} & 0 \\[0.1cm]
-\frac{\partial N_i}{\partial x} & -\frac{\partial N_i}{\partial x} & 0 \\[0.1cm]
-\frac{\partial N_i}{\partial y} & -\frac{\partial N_i}{\partial y} & 0 \\[0.1cm]
0 & 0 & \frac{\partial N_i}{\partial x} \\[0.1cm]
0 & 0 & \frac{\partial N_i}{\partial y} 
\end{bmatrix}.
\end{equation}

\subsection{Stiffness matrix components}

The stiffness matrix is constructed by differentiating the residuals with respect to the nodal variables. The entries related purely to the displacement field may thus be found as: 
\begin{equation}
\bm{K}^{\mathbf{u},\mathbf{u}}_{ij} = \dfrac{\partial \mathbf{R}_i^\mathbf{u}}{\partial \mathbf{u}_j} = \int_\Omega \left[\left(1-\phi\right)^2 + k \right]\left(\bm{B}^\mathbf{u}_i\right)^T\bm{\mathcal{L}}_0\bm{B}^\mathbf{u}_i\, \tm{d}V.
\end{equation}
In a similar manner, the plastic strain field stiffness is given by: 
\begin{equation}
\begin{split}
\bm{K}^{\bm{\varepsilon}^p,\bm{\varepsilon}^p}_{ij} = \dfrac{\partial \mathbf{R}_i^{\bm{\varepsilon}^p}}{\partial \bm{\varepsilon}^p_j} =& \int_\Omega \left\{ \left(\bm{N}^{\bm{\varepsilon}^p}_i\right)^T\left[\left(\dfrac{\partial \bm{q}^D}{\partial\varepsilon^p_j} - \left[\left(1-\phi\right)^2 + k \right]\bm{\mathcal{L}}_0\right)\bm{N}^{\bm{\varepsilon}^p}_j+\dfrac{\partial \bm{q}^D}{\partial\nabla\varepsilon^p_j}\bm{B}^{\bm{\varepsilon}^p}_j\right]  \right. \\ 
&\left. + \left(\bm{B}^{\bm{\varepsilon}^p}_i\right)^T\left(\dfrac{\partial\bm{\tau}^D}{\partial \varepsilon^p_j}\bm{N}^{\bm{\varepsilon}^p}_j + \dfrac{\partial\bm{\tau}^D}{\partial \nabla\varepsilon^p_j}\bm{B}^{\bm{\varepsilon}^p}_j + \dfrac{\partial\bm{\tau}^E}{\partial\nabla\varepsilon^p_j}\bm{B}^{\bm{\varepsilon}^p}_j\right)\right\} \, \tm{d}V.
\end{split}
\end{equation}
The coupling terms are found as: 
\begin{align}
\bm{K}^{\mathbf{u},\bm{\varepsilon}^p}_{ij} = \dfrac{\partial \mathbf{R}_i^\mathbf{u}}{\partial \bm{\varepsilon}^p_j} = -\int_\Omega \left[\left(1-\phi\right)^2 + k \right]\left(\bm{B}^\mathbf{u}_i\right)^T\bm{\mathcal{L}}_0 \bm{N}^{\bm{\varepsilon}^p}_j \, \tm{d}V, \\
\bm{K}^{\bm{\varepsilon}^p,\mathbf{u}}_{ij} = \dfrac{\partial \mathbf{R}_i^{\bm{\varepsilon}^p}}{\partial \mathbf{u}_j} = -\int_\Omega \left[\left(1-\phi\right)^2 + k \right]\left(\bm{N}^{\bm{\varepsilon}^p}\right)^T\bm{\mathcal{L}}_0 \bm{B}^{u}_j\, \tm{d}V.
\end{align}
The stiffness related to the crack phase field is determined in an equivalent manner: 
\begin{equation}
\bm{K}^{\phi,\phi}_{ij} = \dfrac{\partial R^\phi_i}{\partial\phi_j} = \int_\Omega \left[\left(2H+\dfrac{G_c}{\ell}\right)N_iN_j + G_c\ell\mathbf{B}_i^T\mathbf{B}_j\right]\, \tm{d}V.
\end{equation}
Finally, the contributions from the mass transport of hydrogen in the metal lattice include a diffusivity matrix:
\begin{equation}
\bm{K}^{C,C}_{ij} = \int_\Omega\left(\mathbf{B}_i^T\mathbf{B}_j - \mathbf{B}_i^T\dfrac{\overline{V}_H}{RT}\nabla\sigma_H N_j\right)\,\tm{d}V,
\end{equation}
and a concentration capacity matrix:
\begin{equation}
\bm{M}_{ij} = \int_\Omega N_i^T \dfrac{1}{D}N_j\,\tm{d}V.
\end{equation}

The complete element assemble is given by (\ref{eq:CompleteSystem}), where the time derivative of the hydrogen concentration $\dot{C}$ is discretised in an analogous manner to $C$.

\bibliographystyle{elsarticle-harv}
\bibliography{library}


\end{document}